%% file: main.tex
\documentclass{article}

\usepackage{nac_preprint}
\usepackage[utf8]{inputenc} % allow utf-8 input
\usepackage[T1]{fontenc}    % use 8-bit T1 fonts
\usepackage{hyperref}       % hyperlinks
\usepackage{url}            % simple URL typesetting
\usepackage{booktabs}       % professional-quality tables
\usepackage{amsfonts}       % blackboard math symbols
\usepackage{nicefrac}       % compact symbols for 1/2, etc.
\usepackage{microtype}      % microtypography
\usepackage{comment}

\usepackage{booktabs} % For formal tables
\usepackage{subcaption}
\usepackage{amsmath}
\usepackage{amssymb}
\usepackage{amsthm}
\usepackage{cleveref}
\usepackage{mathtools}
\usepackage[toc,page]{appendix}
\usepackage{enumitem}
\usepackage{graphics}
\usepackage{graphicx}
\usepackage{xcolor}
\usepackage[export]{adjustbox}
\usepackage{algorithm}
\usepackage{algorithmicx}
\usepackage[noend]{algpseudocode}
\usepackage{enumitem}
\algnewcommand{\LineComment}[1]{\State \(//\) #1}
\algnewcommand{\RLineComment}[1]{\State \(\triangleright\) #1}
\usepackage{bm}
\usepackage{multirow}

\usepackage{wrapfig}

\newcommand*\samethanks[1][\value{footnote}]{\footnotemark[#1]}

\usepackage{etoolbox}
\usepackage{tikz}
\usetikzlibrary{tikzmark}
\usetikzlibrary{calc}

\errorcontextlines\maxdimen

\newcommand{\ALGtikzmarkcolor}{black}% customise this, if you want
\newcommand{\ALGtikzmarkextraindent}{4pt}% customise this, if you want
\newcommand{\ALGtikzmarkverticaloffsetstart}{-.5ex}% customise this, if you want
\newcommand{\ALGtikzmarkverticaloffsetend}{-.5ex}% customise this, if you want
\makeatletter
\newcounter{ALG@tikzmark@tempcnta}

\newcommand\ALG@tikzmark@start{%
    \global\let\ALG@tikzmark@last\ALG@tikzmark@starttext%
    \expandafter\edef\csname ALG@tikzmark@\theALG@nested\endcsname{\theALG@tikzmark@tempcnta}%
    \tikzmark{ALG@tikzmark@start@\csname ALG@tikzmark@\theALG@nested\endcsname}%
    \addtocounter{ALG@tikzmark@tempcnta}{1}%
}

\def\ALG@tikzmark@starttext{start}
\newcommand\ALG@tikzmark@end{%
    \ifx\ALG@tikzmark@last\ALG@tikzmark@starttext
    \else
        \tikzmark{ALG@tikzmark@end@\csname ALG@tikzmark@\theALG@nested\endcsname}%
        \tikz[overlay,remember picture] \draw[\ALGtikzmarkcolor] let \p{S}=($(pic cs:ALG@tikzmark@start@\csname ALG@tikzmark@\theALG@nested\endcsname)+(\ALGtikzmarkextraindent,\ALGtikzmarkverticaloffsetstart)$), \p{E}=($(pic cs:ALG@tikzmark@end@\csname ALG@tikzmark@\theALG@nested\endcsname)+(\ALGtikzmarkextraindent,\ALGtikzmarkverticaloffsetend)$) in (\x{S},\y{S})--(\x{S},\y{E});%
    \fi
    \gdef\ALG@tikzmark@last{end}%
}

\apptocmd{\ALG@beginblock}{\ALG@tikzmark@start}{}{\errmessage{failed to patch}}
\pretocmd{\ALG@endblock}{\ALG@tikzmark@end}{}{\errmessage{failed to patch}}
\makeatother
\newtheorem{theorem}{Theorem}
\newtheorem{corollary}{Corollary}[theorem]

\title{Time-Integrated Spike-Timing-Dependent-Plasticity }

\author{%
William Gebhardt\samethanks\\
Rochester Institute of Technology \\ 
\texttt{wdg1351@rit.edu}
\And
Alexander G. Ororbia\samethanks \\
Rochester Institute of Technology \\
\texttt{ago@cs.rit.edu}
}

\begin{document}

\setlength{\abovedisplayskip}{0.065cm}
\setlength{\belowdisplayskip}{0pt}

\maketitle

\begin{abstract} 
In this work, we propose time-integrated spike-timing-dependent plasticity (TI-STDP), a mathematical model of synaptic plasticity that allows spiking neural networks to continuously adapt to sensory input streams in an unsupervised fashion. Notably, we theoretically establish and formally prove key properties related to the synaptic adjustment mechanics that underwrite TI-STDP. 
%TI-STDP lies in the middle of a spectrum of STDP processes, i.e. in between classical STDP and event-driven STDP, and is able to recover and subsume the dynamics elicited by the schemes at the extreme ends of this computational modeling continuum. 
Empirically, we demonstrate the efficacy of TI-STDP in simulations of jointly learning deeper spiking neural networks that process input digit pixel patterns, at both full image and patch-levels, comparing to two powerful historical instantations of STDP; trace-based STDP (TR-STDP) and event-based post-synaptic STDP (EV-STDP). Usefully, we demonstrate that not only are all forms of STDP capable of meaningfully adapting the synaptic efficacies of a multi-layer biophysical architecture, but that TI-STDP is notably able to do so without requiring the tracking of a large window of pre- and post-synaptic spike timings, the  maintenance of additional parameterized traces, or the restriction of synaptic plasticity changes to occur within very narrow windows of time. This means that our findings show that TI-STDP can efficiently embody the benefits of models such as canonical STDP, TR-STDP, and EV-STDP without their costs or drawbacks. Usefully, our results further demonstrate the promise of using a spike-correlation scheme such as TI-STDP in conducting credit assignment in discrete pulse-based neuromorphic models, particularly those than acquire a lower-level distributed representation jointly with an upper-level, more abstract representation that self-organizes to cluster based on inherent cross-pattern similarities. We further demonstrate TI-STDP's effectiveness in adapting a simple neuronal circuit that learns a simple bi-level, part-whole hierarchy from sensory input patterns.
% (without using label information except for post-simulation analysis).

\end{abstract}

\section{Introduction}
\label{sec:intro}

Among its many objectives, brain-inspired computing seeks to develop effective neuromimetic architectures as well as their underpinning learning or credit assignment algorithms \cite{ororbia2023brain, ororbia2024review}, particularly those that would prove useful when implemented on in-memory computing hardware \cite{mead1990neuromorphic, draghici2000neural, furber2016large, massa2020efficient, zhang2021optical}. In particular, spiking neural networks (SNNs), the third generation of artificial neural networks (ANNs) \cite{maass1997networks}, have emerged as a class of systems that exhibit properties that circumvent many of the key limitations facing backprop-based ANNs, including their long-criticized biological implausibility \cite{grossberg1987resonance, crick1989recent}, energy inefficiency, as well as their difficulty in generalizing to data streams with fewer resources \cite{ororbia2023spiking, ororbia2023learning} using sparse computations. 

Nevertheless, despite their promise, there are challenges that arise when working with SNNs and spike-based computation. Primarily, there is much contention as to how to properly adjust a spiking network's synaptic plasticity. Put in another manner, a long-standing, central question that has driven and motivated many efforts in brain-inspired computing and computational neuroscience is: what is an effective, efficient means of conducting credit assignment in a biophysical architectures such as networks composed of spiking neuronal units? 
In attempting to answer this important question, there have been many proposed methods, ranging from approximating gradients \cite{bellec2020solution}, to Hebbian plasticity \cite{hebb1949organization} and Hebbian-based methods \cite{gupta2009hebbian, samadi2017deep}, to the method that we will be focusing on in this work -- spike-timing-dependent-plasticity (STDP) \cite{markram1997physiology, markram1997regulation, bi2001synaptic}. Spike-timing-centered forms of plasticity are particularly interesting and important given that, generally, the only information that is required to perform a synaptic update is that related to  locally-available (in terms of both model structure/space) statistics related to neuronal spike emissions. The canonical, complete form of STDP generally involves utilizing all, or a reasonably-sized window, of spike firing times encountered during the presentation of a sensory stimulus signal, making it prohibitively costly to simulate whereas fast, online approximations to it involve maintaining experimenter-tuned, parametric trace variables \cite{diehl2015unsupervised} or making particular assumptions/simplifications about how to engage in spike-triggered synaptic alteration \cite{tavanaei2018representation} (furthermore, rarely are these different forms of STDP cross-examined in the same setting). % AO: maybe? not sure about this statement just yet...

In this work, we propose a novel method for plasticity adjustment based solely on the pre- and post-synaptic spike timings, without requiring the maintenance of any auxiliary variable traces or controlled explicit windows of (all or a large swath of) spike timings. We will show how this method aligns itself to the input data to effectively extract a diversity of visual receptive fields as well as its efficacy in adapting the parameters of a multi-layer spiking network that must jointly acquire several self-organizing levels of sensory input representation. Empirically, our model of spike-timing driven synaptic adjustment is compared to several important models of STDP, where we measure their utility in adapting an unsupervised model that is directly utilized for classification. In addition, we construct a TI-STDP-driven compositional spiking neuronal circuit that is tasked with learning a bi-level, part-whole hierarchy from raw sensory information, where we further introduce a simple ancestral assembly process for synthesizing object patterns from the knowledge encoded in the model's synaptic efficacies.

%% AO: i think this is getting implicity covered at the start of the next section, technically...
\begin{comment} 
\section{Related Work}
\label{sec:lit_review}

\textbf{TODO:} \textcolor{red}{Write This: } 
Here, your paper should introduce the current methods of solving the problem you are studying and should include most, if not all, of the methods used in the benchmarks (these will likely even be some of your implemented baselines that you will be comparing to / against). It is important to do a thorough job incorporating whatever background knowledge you wish the reader to have here before they continue reading on to your proposed method/idea/algorithm/theorem. Length will vary depending on your target (conferences - this section will be somewhat shorter, journals - this section will be substantial/detailed). Citations are important XXX.
\end{comment}

\section{Methodology}
\label{sec:method}

In this section, we begin by describing the neuronal dynamics of the spiking units that we investigate in this work, followed by our implementations of related key forms of spike-timing-dependent plasticity. We then turn our attention to presenting our proposed time-integrated spike-timing-dependent plasticity and theoretically prove several of its key properties. A general overview of the similarities and differences between all the methods can be found in Table \ref{tab:stdp_similarities}.
% and further mathematically demonstrate its equivalence with canonical spike-timing-dependent plasticity and (post-synaptic) event-based spike-timing-dependent plasticity.

\subsection{Neural Architecture and Spiking Dynamics}

\paragraph{Neuronal Structure and Dynamics} 
In this study, we construct neuronal systems composed of leaky integrate-and-fire (LIF) neurons. The equations for these biomemetic neurons were first proposed by Stein \cite{stein1965theoretical} in 1965. Since then the dynamics for an LIF's voltage ($v^\ell_k$) have been distilled to the following ordinary differential equation (ODE): 
% and its spike emission ($s^\ell_k$):
\begin{align}
    \tau_m \frac{\partial v^\ell_k(t)}{\partial t} = -\gamma_v v^\ell_k(t) + R^\ell j^\ell_k(t)
\end{align}
% tau, gamma, R
where $\tau_m$ is the membrane potential time constant, $\gamma_v$ is the voltage leak coefficient, and $R^\ell$ is the membrane resistance. For this paper when calculating the the LIF's incoming electrical current $j^\ell_k(t)$, we treat the electrical current as a point wise current, reducing the time a signal is active to a single pass. To calculate this point wise current we treat the current as simply the weighted sum of incoming signals: 
$j^\ell_k(t) = \sum_i W^\ell_{ki} s^{\ell-1}_i$ which becomes equivalent to the sum of weight values for non-zero pre-synaptic spike variables, i.e., 
\begin{align}
    j^\ell_k(t) = \sum_i
    \begin{cases}
        W^\ell_{ik} & s^{\ell-1}_i = 1\\
        0 & \text{otherwise}.
    \end{cases} 
\end{align}
where $s^{\ell-1}_i$ is a binary value denoting if there is a spike emission from presynaptic neuron $i$. The equation governing the spike emissions ($s^\ell_k)$ for a given LIF neuron $k$ is as follows
\begin{align}
    s^\ell_k(t) = 
    \begin{cases}
        1 & v^\ell_k(t) > \theta^\ell_k(t) \\
        0 & \text{otherwise}
    \end{cases} 
\end{align}
where $\theta^\ell_k(t)$ is the membrane voltage threshold (which, once breached, results in the neuron emitting a discrete spike pulse). 
The voltage threshold is notably a function of time and is constructed to adhere to its own dynamics (making it an `adaptive threshold'). Specifically, the threshold is a combination of a time-varying homeostatic variable $\hat{\theta}^\ell_k(t)$ and a fixed base threshold value $v^\ell_{base}$ as follows:
\begin{align}
    \frac{\partial \hat{\theta}^\ell_k(t)}{\partial t} = -\frac{1}{\tau_\theta}\hat{\theta}^\ell_k(t) + \kappa_\theta s^\ell_k(t), \quad \theta^\ell_k(t) = v^\ell_{base} + \hat{\theta}^\ell_k(t)
\end{align}
where $\tau_\theta$ is the homeostatic variable time constant and $\kappa_\theta$ is the variable increment value.

%\paragraph{Neuronal Architecture} \textcolor{red}{WRITEME:}

%% STDP adaptation / learning
\subsection{Adaptation through Varieties of Spike-Timing-Dependent-Plasticity} 

Spike-timing-dependent plasticity has a reasonably long history in the domain of computational neuroscience and the modeling of event-driven neurons that emit action potentials \cite{bi1998synaptic, markram2011history}. Crucially, STDP as model of (experience-dependent, long-term) learning is viewed as strongly biologically-plausible, given the steady accumulation of neuro-physiological evidence that has been gathered in support of its place in natural neuronal networks. The rule was initially observed by Bi and Poo, and markram in their studies of neuronal systems in rats. In the study by Bi and Poo they observed that by stimulating a pair of neurons in certain sequences they could adjust the synaptic plasticity of the axon connecting the pair. 
%Shocking a rat

\paragraph{Canonical Pairwise STDP} In canonical STDP \cite{gerstner1996neuronal,kempter1999hebbian}, a synaptic connection's strength is changed as a function of the relative timing between pre-synaptic spikes and post-synaptic spikes; in effect, unlike the well-known phrase ``neurons that fire together, wire together'' (which pithily summarizes the ideas of \cite{hebb1949organization}),  neuronal cells that fire together do not necessarily always wire together given that the timing of their firing matters as well. If we label a pre-synaptic neuron's spike emission time as $t^{\ell-1}_i$ and a post-synaptic neuron's spike emission time as $t^{\ell}_j$, then the instantaneous adjustment induced is:
\begin{align}
\label{eqn:classical_stdp}
    \Delta W^\ell_{ij} = 
    \begin{cases}
        A_{+} e^{\frac{-(t^\ell_j - t^{\ell-1}_i)}{\tau_{+}}} & t^\ell_j > t^{\ell-1}_i \\
        - A_{-} e^{\frac{t^\ell_j - t^{\ell-1}_i}{\tau_{-}}} & t^\ell_j < t^{\ell-1}_i \\ 
        0 & t^\ell_j = t^{\ell-1}_i \\
    \end{cases} 
\end{align}
where $A_{+}$ controls the strength of the long-term potentiation (LTP) applied to synapse $W^\ell_{ij}$ and $A_{-}$ mediates the degree to which long-term depression (LTD) is applied. This method triggers whenever there is a new pre or post synaptic spike.

%% total synaptic weight change
\begin{comment}
Formally, over a stimulus presentation time (or window length) $T$ milliseconds (ms), if we tag any specific pre-synaptic spike time as $t^{\ell-1}_{j,p}$ and label any particular post-synaptic time as $t^\ell_{i,q}$, then classical STDP (fit to experimental data in neuroscience studies) amounts to:
\begin{align}
\Delta W^\ell_{ij} = \sum^{T^{\ell-1}_p}_{p=1} \sum^{T^\ell_q}_{q=1}  
\begin{cases} 
A_{+} \exp(-(t^\ell_{i,q} - t^{\ell-1}_{j,p})) & t^\ell_{i,q} - t^{\ell-1}_{j,p} > 0 \\ 
-A_{-} \exp(-(t^\ell_{i,q} - t^{\ell-1}_{j,p})) & t^\ell_{i,q} - t^{\ell-1}_{j,p} < 0 \label{eqn:classical_stdp}
\end{cases}
\end{align}
where $T^{\ell-1}_p = \{t^{\ell-1}_{j,1}, t^{\ell-1}_{j,2},...\}$ is the set of recorded pre-synaptic firing times and $T^\ell_q = \{t^\ell_{i,1}, t^\ell_{i,2},...\}$ is the set of recorded post-synaptic firing times. $A_{+}$ controls the strength of the long-term potentiation (LTP) applied to synapse $W^\ell_{ij}$ where $A_{-}$ mediates the degree to which long-term depression (LTD) is applied. Variations of the timing-centric of STDP sometimes choose to employ a moving window of spike timings to reduce the computational burden posed by the original rule's examination of all pre- and post-synaptic firing time tags within a stimulus window.
\end{comment}

\paragraph{Trace-Based STDP (TR-STDP), with Pre-Synaptic Disconnect} %The canonical form of STDP presented above quickly becomes prohibitively expensive as the length of time it is applied increases. Beyond truncating the length of the window, another way to implement STDP learning is via traces of neuronal activity dynamics. 
One way of instantiating the above form of STDP is through the introduction of variable traces that  decay exponentially with time. The biophysical interpretation of these traces is that they represent the concentrations of particular molecules or ions\cite{carafoli1987intracellular,karmarkar2002mechanisms,huang2004glutamate}\footnote{For a pre-synaptic neuron, a trace can be biologically interpreted as the fraction of activated NMDA receptors (those that are in the `open state') \cite{huang2004glutamate}. In contrast, for a post-synaptic neuron, a trace variable value can be viewed as the calcium concentration in the dendritic spine of the post-synaptic neuron (deposited when a backpropagating action potential 
reaches this postsynaptic spine) \cite{carafoli1987intracellular}.} and decrease or increase with time in the presence of events (such as spike pulses); the magnitude/level of these traces effectively serve to temporally code sequences of events. 
Formally, the trace $z^\ell_k(t)$ of the output spike $s^\ell_k(t)$ of the $k$th neuronal cell in layer $\ell$ follows either the ordinary differential equation of the form as shown below:
\begin{align}
   \tau_z \frac{\partial z^\ell_k(t)}{
   \partial t} = 
   -z^\ell + \gamma_z s^\ell_k(t) \label{eqn:full_trace}
\end{align}
where $\gamma_z$ controls the amount incrementally added to  cells' trace values. Alternatively, a trace vector can be maintained with the simple piece-wise (binary-switch) update equation:
\begin{align}
    z^\ell_k(t + \Delta t) = s^\ell_k(t) + \Big( z^\ell_k(t) - \frac{\Delta t}{\tau_{z}} z^\ell_k(t) \Big) (1 - s^\ell_k(t)) \label{eqn:switch_trace}
\end{align}
which effectively sets a cell $k$'s trace value to $1$ there is spike activity and otherwise the current trace value decays exponentially. We found the second form (Equation \ref{eqn:switch_trace}) to work best for this study's experiments.

Given the variable trace above, a synaptic efficacy $W^\ell_{ij}$ between pre-synaptic neuron $i$ and post-synaptic neuron $j$ may be adjusted as follows:
\begin{align}
    \Delta W^\ell_{ij} = A_+ \overbrace{ \Big(1 - W^\ell_{ij} \Big)^{\mu} \Big( \big( \mathbf{z}^{\ell-1}_i(t) - \mathbf{z}^{\ell-1}_{tar} \big) \big( \mathbf{s}^\ell_j(t) \big)^T \Big)}^{\text{Long-term Potentiation}} - A_{-} \overbrace{\Big( W^\ell_{ij} \Big)^{\mu} \Big( \big( \mathbf{s}^{\ell-1}_i(t) \big)  \big( \mathbf{z}^\ell_j(t) \big)^T \Big)}^{\text{Long-term Depression}} \label{eqn:trace_stdp}
\end{align}
where $A_+$ is the coefficient that controls the strength of the potentiation and $A_{-}$ is the coefficient controlling the strength of the depression applied. $\mu$ controls the effect that the weight dependency has on the Hebbian adjustment applied to synapses (set in our experiments to $\mu = 1$).

Note that in the plasticity dynamics above, $z^{\ell-1}_{tar}$ contains target trace value(s); when this is set to a value greater than zero, a form of pre-synaptic `disconnect' is integrated into the STDP updates. For any layer $\ell$, a non-negative value of $z^{\ell-1}_{tar}$ provides an approximate target activity level for each neuron in $\ell-1$ to reach -- the higher this value, the lower the synaptic strengths will be and the more that pre-synaptic neurons that spike infrequently will be ``pruned away'' \cite{diehl2015unsupervised} over time as synaptic changes are applied. 
Note that introducing a non-negative $z^{\ell-1}_{tar}$ will result in negative synaptic updates (if the pre-synaptic trace has decayed to values below $z^{\ell-1}_{tar}$) -- this has the desirable effect of disconnecting pre-synaptic neurons that have little to no effect in causing post-synaptic spike events (yielding a simple form of weight decay, endowing a degree of noise robustness to the post-synaptic neuron's integration of incoming signals). 
Notice that Equation \ref{eqn:trace_stdp} presents the full trace-based from of STDP above, making explicit both the post-synaptic (weighted by $A_+$) and pre-synaptic (weighted by $A_{-}$) terms as well as the weight-dependency that is integrated. 

\paragraph{Event-Based Post-Synaptic STDP (EV-STDP)} Lying on the other extreme of STDP formulations are those that operate with particular events, rather than operating with all previously encountered pre- and post-synaptic spike firing times. In \cite{tavanaei2018representation}, such a form of STDP was proposed, which focused on using post-synaptic spikes to trigger synaptic adjustment that was a function of the synaptic weight values and the presence of pre-synaptic spikes within a small window of time. As this form of STDP lies on a particularly interesting end of the STDP form spectrum, we implement and investigate an instantiation in \cite{tavanaei2018representation} that we will refer to as event-based STDP (EV-STDP).

The synaptic update induced by EV-STDP, reformulated to adhere to the mathematical context and framing of this paper, is formally the following: 
\begin{align}
    \Delta W^\ell_{ij}(t) = A_{+} \overbrace{ \Big( (1 - W^\ell_{ij} (1 + \lambda)) f^{\ell-1}_i(t)  \Big) s^\ell_j(t) }^{\text{Long-term Potentiation}} + 
    A_{-} \overbrace{ \Big( -W^\ell_{ij} (1 + \lambda) (1 - f^{\ell-1}_i(t))  \Big) s^\ell_j(t) }^{\text{Long-term Depression}} \label{eqn:event_stdp}
\end{align}
% \begin{align}
%     \Delta W^\ell_{ij}(t) = 
%     \begin{cases} 
%       A_{+} (1 - W^\ell_{ij}) (1 + \lambda) s^\ell_i & s^{\ell-1}_j = 1 \\
%       A_{-} -W^\ell_{ij} (1 + \lambda) s^\ell_i & s^{\ell-1}_j = 0
%    \end{cases}
% \end{align}
where $f^{\ell-1}_i(t)$ checks for the presence of a spike within a small, controlled time window of $[t - t_\epsilon, t]$ -- this function emits a value of one if a spike happened between $t - t_\epsilon$ and $t$ milliseconds (and zero otherwise); we set $t_\epsilon = 1$ ms. 
In the above STDP update, we see that only the post-synaptic spikes (within a small window of time) matter in triggering the potentiation and depression pressures produced by this scheme. Note that, in EV-STDP, $\lambda$ is a scalar factor that controls the degree to which synaptic efficacies undergo LTP and LTD.

\begin{table}[!t]
    \centering
    \begin{tabular}{|c|c c c | c c c | c | c|}
    \hline
        & \multicolumn{3}{c|}{\textbf{Pre-synaptic}} & \multicolumn{3}{c|}{\textbf{Post-synaptic}} & \textbf{Simulation} & \textbf{Weight} \\
        \textbf{Model} & \textbf{Event} & \textbf{Spike-Time} & \textbf{Trace} & \textbf{Event} & \textbf{Spike-Time} & \textbf{Trace} & \textbf{Time} & \textbf{Dependent} \\ \hline
        STDP & \checkmark & \checkmark & & \checkmark & \checkmark & & & \\ \hline
        TR-STDP & \checkmark & & \checkmark & \checkmark & & \checkmark & & \checkmark \\ \hline
        EV-STDP & \checkmark & & \checkmark & \checkmark & & \checkmark & & \checkmark \\ \hline
        TI-STDP & & \checkmark & & & \checkmark & & \checkmark & \checkmark \\ \hline
    \end{tabular}
    \vspace{0.2cm}
    \caption{\textbf{Plasticity Model Conceptual Comparison.} In this table, we highlights the needs of various spike-timing plasticity models with respect to what aspects of temporal information they utilize such as: pre-synaptic activity, post-synaptic activity, the current simulation time, and, finally, the whether or not the form of synaptic plasticity is made weight-dependent.}
    \label{tab:stdp_similarities}
    \vspace{-0.4cm}
\end{table}
% \begin{figure}[!t]
%     \centering
%     \includegraphics[width=0.8\textwidth]{figs/stdp_hierarchy.pdf} 
%     \caption{The conceptual hierarchy underlying spike-timing-dependent plasticity (STDP) and time-integrated STDP's place within this concept-relation hierarchy.} \label{fig:stdp_hierarchy} % maybe make a table outlining differences of tistdp and others?
% \end{figure}

\subsection{Time-Integrated Spike-Timing-Dependent-Plasticity} 
\label{sec:TISTDP}

In the context of spike-timing-based adaption, we propose time-integrated spike-timing-dependent plasticity (TI-STDP) for modulating an SNN's synaptic efficacies, which is based on a three term/variable system. This form of synaptic adjustment is conducted exclusively with locally available temporal information -- the time $t_i$ of the last presynaptic spike $s_i$, the time $t_j$ of the last postsynaptic spike $s_j$, and the current time $t$. From these three terms, we may then calculate an incremental update at each and every time step, making TI-STDP naturally an online synaptic evolution process. In this section, we will first describe each of the core mechanisms (and their relevant spiking conditions) that underwrite TI-STDP and then put all of them together to present the full plasticity dynamics induced by the proposed rule.

\paragraph{Post-Synaptic Event-Based Scale Factor} In TI-STDP, we integrate an exponentially decaying scale factor based on the current time $t$ and the time of the last post-synaptic event/time $t_j$. The driving motivation behind this term is to reduce/mitigate the continued integration that is caused by post-synaptic events in the past. In effect, our mathematical model emphasizes the fact that this post-synaptic-driven term needs to function the same regardless of the timescale that the neuronal dynamics are operating on; the reason for this design choice is that, if the timescale of the model is $dt=100$ ms versus $dt=1$ ms, a single time step difference would induce far too strong of a synaptic decay. For this reason, this part of the synaptic dynamics decays as a function of simulation steps and not actual times. As a result, the general form of this mechanism becomes: $e^{(t_j - t) / \Delta t}$. Note that, as $t_j \leq t$ for all points in time, the largest that this value will ever be is $1$ and it will decay down to $0$ as $t_j - t$ approaches $-\infty$.

\paragraph{Silent Pre-Synaptic Neuron} In the event that there is this a post-synaptic event and a pre-synaptic neuron has been `silent' (it has not spiked for a long period of time) the plasticity between these two neurons decays. This decay is specifically constructed to be a constant factor that is scaled by the post-synaptic scale term. This decay factor $\gamma$ is usually chosen to be between $0$ and $1$ for optimal results, i.e., $\gamma \in [0,1]$. The resulting equation for the decay, which is further integrated with a weight dependency, is as follows:
\begin{equation}
\label{eqn:quietTI}    
    \frac{dW^\ell_{ij}(t)}{dt} = -\gamma * e^{(t_j - t) / \Delta t} * W^\ell_{ij}(t).
\end{equation}
Notice that the decay is scaled by the current synaptic efficacy $W^\ell_{ij}$ so as to cause the decay to be stronger in neurons that have a larger synaptic efficacy and smaller in the synaptic weights with values closer to zero.

\paragraph{Noisy Pre-Synaptic Neuron} In the case that there is a signal transmitted from the presynaptic neuron, the plasticity of the synapse between the pre- and post-synaptic neuron is adjusted based on the spike-timing of each. Both of these cases are wrapped up in the dynamics governed by the following equation:
\begin{equation}
\label{eqn:noisyTI}
    \frac{dW^\ell_{ij}(t)}{dt} = \frac{-\beta}{(t_i - t_j) / \Delta t - 0.5} * e^{(t_j - t) / \Delta t} * (1 - W^\ell_{ij}(t))
\end{equation}
where $\beta$ is a non-negative control factor used to modulate the resulting synaptic adjustment. We notice that this equation is quite similar to the decay term described for the earlier case of the silent pre-synaptic neuron; however, the difference between the pre- and post-synaptic activity scales the update down if there is a large gap between the (pre- and post-synaptic) spike times. Note that this divisor is further shifted by the addition of a constant $-0.5$ term; this represents the fact that, in the case that in a single propagation of information across synapse $W^\ell_{ij}$, if both the pre- and post-synaptic neurons spike, there is half of a time-step difference between the layers of neurons firing. This is scaled by a reverse weight dependency $1-W^\ell(t)$ term so as to cause the updates to be stronger if the spike pattern is abnormal and weaker if it is more common.

\begin{figure*}[!t]
     \centering
     \begin{subfigure}[b]{0.485\textwidth}
         \centering
         \includegraphics[width=1.0\textwidth]{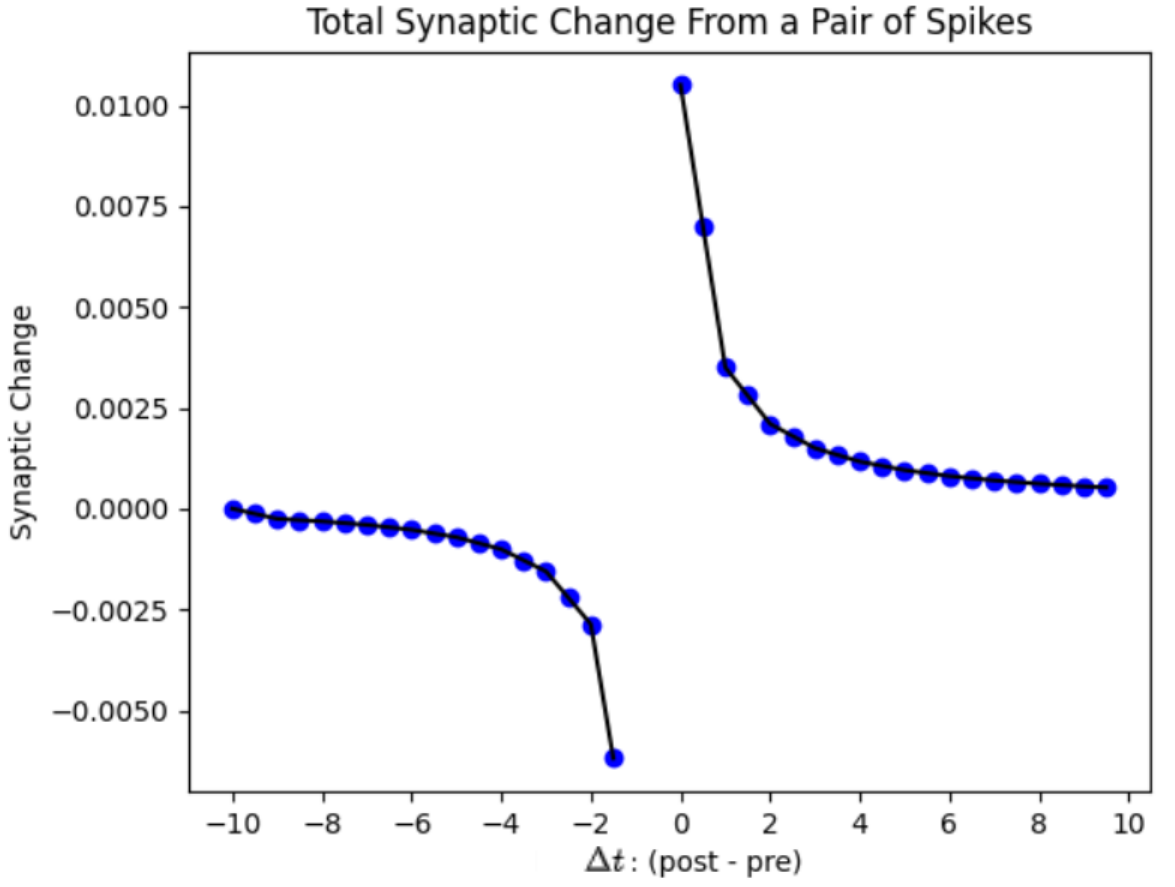}
         \caption{Our TI-STDP process.}
         \label{fig:tistdp_shape}
     \end{subfigure}
     %\hspace{0.05cm}
     \begin{subfigure}[b]{0.485\textwidth}
         \centering
         \includegraphics[width=0.9\textwidth]{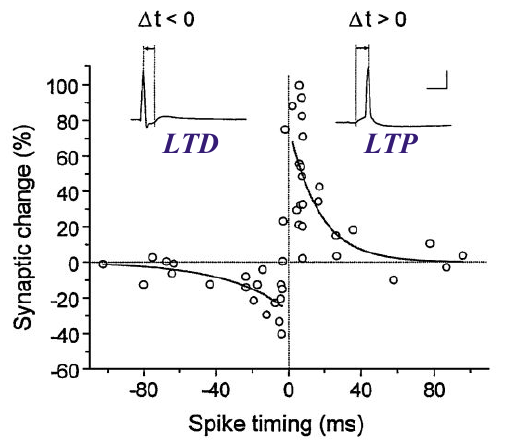}
         \caption{Empirical STDP (plot adapted from \cite{bi2001synaptic}).}
         \label{fig:stdp_shape}
    \end{subfigure}
        \caption{Functional plot of synaptic efficacy adjustments produced by the proposed TI-STDP (Left). The produced synaptic updates by our process closely align with the experimentally-derived STDP (Right), produced by \cite{bi2001synaptic}.}
        \label{fig:shape_of_stdp}
        \vspace{-0.5cm}
\end{figure*}

In Figure \ref{fig:shape_of_stdp}, we visualize the functional shape of the synaptic adjustments produced by the core dynamics of TI-SDP in comparison to the classical curve yielded by (canonical) STDP, notably as it has been empirically-derived and corraborated by a variety experiments conducted history \cite{kempter1999hebbian, markram1997physiology,gerstner1996neuronal, bi1998synaptic,bi2001synaptic}.

\paragraph{TI-STDP Accumulation of Deltas}
Unlike the previous methods, TI-STDP accumulates the update induced by a pair of spikes in time until there is another update. In order to compare the updates between different STDP-like learning rules, TI-STDP must be solved for the total update caused by a pair of spikes. Theorem \ref{thr:1} shows the relation between the current time $t$ and the synaptic weight values at the time of the last encountered (spike) activity. Note that since this theorem considers a single spike in both the pre- and post-synaptic layer, the addition of a new spike would cause the dynamics to change and thus a new $W(t_l)$ would be used.\footnote{Note that, for clarity and simplicity in presenting our theorem, without any loss of generality, we omit the layer index superscript $\ell$ and the pre- and post-synaptic neuron indices $i$ and $j$. The theorem is independent of the physical location of the synapse and neurons within an architecture and specifically applies to the three-variable system composed of a single pre-synaptic neuronal cell connected to a single post-synaptic neuronal cell.} The TI-STDP weight change theorem can formally be stated as follows:
\begin{theorem}
    \label{thr:1}
    Given the existence of activity in both the pre-synaptic neuron ($i$) and post-synaptic neuron ($j$), the equation governing the synaptic plasticity at a given time $t$ with a known $W(t_l)$ is:
    \begin{equation}
        W(t) = 1 + (W(t_l) - 1)e^{\frac{\beta}{(t_i - t_j) / \Delta t - 0.5}(e^{(t_j - t_l) / \Delta t} - e^{(t_j - t) / \Delta t})}
    \end{equation}
    \begin{proof}
        Let $t_i$ and $t_j$ be the pre- and post-synaptic spike timings formally represented as positive integers and let $t_l$ represent the latest synaptic event and further let it be equal to the maximum of $t_i$ and $t_j$. Assume that all constants are positive real numbers. It is also important to note that the bounds for $W$ are $0 < W < 1$.
    
        When $t_i > 0$ and $t_j > 0$, the equation for $dW(t)/dt$ becomes:
        \[
            \frac{dW(t)}{dt} = \frac{-\beta}{(t_i - t_j) / \Delta t - 0.5} * e^{(t_j - t) / \Delta t} * (1 - W(t))
        \]
        To solve for $W(t)$ the indefinite integral can be shown by:
        \begin{align*}
            \frac{dW(t)}{dt} &= \frac{-\beta}{(t_i - t_j) / \Delta t - 0.5} * e^{(t_j - t) / \Delta t} * (1 - W(t)) \\
            \frac{dW(t)}{1 - W(t)} &= \frac{-\beta}{(t_i - t_j) / \Delta t - 0.5} * e^{(t_j - t) / \Delta t} dt \\
            \int \frac{dW(t)}{1 - W(t)} &= \int \frac{-\beta}{(t_i - t_j) / \Delta t - 0.5} * e^{(t_j - t) / \Delta t} dt \\
            - ln(|1-W(t)|) + C_1 &= \frac{\beta}{(t_i - t_j) / \Delta t - 0.5} * e^{(t_j - t) / \Delta t} + C_2 \\
            ln(|1-W(t)|) &= \frac{-\beta}{(t_i - t_j) / \Delta t - 0.5} * e^{(t_j - t) / \Delta t} + C \quad \text{Note: } C = C_2 - C_1 \\
            |1-W(t)| &= Ce^{\frac{-\beta}{(t_i - t_j) / \Delta t - 0.5} * e^{(t_j - t) / \Delta t}} \\
            W(t) &=1 + Ce^{\frac{-\beta}{(t_i - t_j) / \Delta t - 0.5} * e^{(t_j - t) / \Delta t}}
        \end{align*}
        To solve for the constant $C$, let $t=t_l$ as this is the synaptic weight efficacies at the start of the synaptic update. Then, solving for $C$ entails the the steps below:
        \begin{align*}
            W(t) &=1 + Ce^{\frac{-\beta}{(t_i - t_j) / \Delta t - 0.5} * e^{(t_j - t) / \Delta t}} \\
            W(t_l) &=1 + Ce^{\frac{-\beta}{(t_i - t_j) / \Delta t - 0.5} * e^{(t_j - t_l) / \Delta t}} \\
            C &= (W(t_l) - 1)e^{\frac{\beta}{(t_i - t_j) / \Delta t - 0.5} * e^{(t_j - t_l) / \Delta t}} \\
            W(t) &= 1 + \Big( (W(t_l) - 1)e^{\frac{\beta}{(t_i - t_j) / \Delta t - 0.5} * e^{(t_j - t_l) / \Delta t}} \Big) e^{\frac{-\beta}{(t_i - t_j) / \Delta t - 0.5} * e^{(t_j - t) / \Delta t}} \\
            W(t) &= 1 + (W(t_l) - 1)e^{\frac{\beta}{(t_i - t_j) / \Delta t - 0.5}(e^{(t_j - t_l) / \Delta t} - e^{(t_j - t) / \Delta t})}
        \end{align*}
    \end{proof}
\end{theorem}
We next formally prove two important corollaries related to the direction of the synaptic efficacy changes that follow from the TI-STDP theorem. Specifically, we consider the case where the non-negative spike-time $t_j$ of the post-synaptic neuron $j$ occurs after the non-negative spike time $t_i$ of the pre-synaptic neuron $i$ and vice versa (where post-synaptic spike time $t_j$ occurs before pre-synaptic spike time $t_i$). 

For the situation that the post-synaptic spike occurs \emph{after} the pre-synaptic spike, which induces a positive shift in synaptic efficacy (or long-term synaptic potentation), we have the following:
\begin{corollary}
Given $0 < t_i \leq t_j$ then $W(t) \geq W(t_j)$.
\begin{proof}
    Let $(t_i - t_j)/\Delta t - 0.5 = -\kappa$, where $\kappa \in \mathbb{R}^+$. Remember that $0 < W(t_j) < 1$. It follows that: 
    \begin{align*}
        W(t) &\geq W(t_j) \\
        1 + (W(t_l) - 1)e^{\frac{\beta}{- \kappa}(e^{(t_j - t_l) / \Delta t} - e^{(t_j - t) / \Delta t})} &\geq W(t_l) \\ 
        (W(t_l) - 1)e^{\frac{\beta}{- \kappa}(e^{(t_j - t_l) / \Delta t} - e^{(t_j - t) / \Delta t})} &\geq W(t_l) - 1 \\
        e^{\frac{\beta}{- \kappa}(e^{(t_j - t_l) / \Delta t} - e^{(t_j - t) / \Delta t})} &\leq 1 \\
        \frac{\beta}{- \kappa}(e^{(t_j - t_l) / \Delta t} - e^{(t_j - t) / \Delta t}) &\leq 0 \\
        e^{(t_j - t_l) / \Delta t} - e^{(t_j - t) / \Delta t} &\geq 0 \\
        e^{(t_j - t_l) / \Delta t} &\geq e^{(t_j - t) / \Delta t} \\
        (t_j - t_l) / \Delta t &\geq (t_j - t) / \Delta t \\
        t_l &\leq t
    \end{align*}
\end{proof}
\end{corollary}

For the situation that the post-synaptic spike occurs \emph{before} the pre-synaptic spike, which induces a negative shift in synaptic efficacy synaptic (or long-term depression), we have the following:
\begin{corollary}
Given $0 < t_j < t_i$ then $W(t) \leq W(t_j)$.
\begin{proof}
    Let $(t_i - t_j)/\Delta t - 0.5 = \kappa$ where $\kappa \in \mathbb{R}^+$. Remember that $0 < W(t_j) < 1$. It follows that:
    \begin{align*}
        W(t) &\leq W(t_j) \\
        1 + (W(t_l) - 1)e^{\frac{\beta}{\kappa}(e^{(t_j - t_l) / \Delta t} - e^{(t_j - t) / \Delta t})} &\leq W(t_l) \\ 
        (W(t_l) - 1)e^{\frac{\beta}{\kappa}(e^{(t_j - t_l) / \Delta t} - e^{(t_j - t) / \Delta t})} &\leq W(t_l) - 1 \\
        e^{\frac{\beta}{\kappa}(e^{(t_j - t_l) / \Delta t} - e^{(t_j - t) / \Delta t})} &\geq 1 \\
        \frac{\beta}{\kappa}(e^{(t_j - t_l) / \Delta t} - e^{(t_j - t) / \Delta t}) &\geq 0 \\
        e^{(t_j - t_l) / \Delta t} - e^{(t_j - t) / \Delta t} &\geq 0 \\
        e^{(t_j - t_l) / \Delta t} &\geq e^{(t_j - t) / \Delta t} \\
        (t_j - t_l) / \Delta t &\geq (t_j - t) / \Delta t \\
        t_l &\leq t        
    \end{align*}
\end{proof}
\end{corollary}
We next turn to formally state and prove TI-STDP's synaptic weight change dynamics in the scenario where these is a post-synaptic spike/event but no pre-synaptic activity.
\begin{theorem}
    \label{thr:2}
    Given the existence of  only postsynaptic activity in neuron $j$ at time $t_j$, the equation that governs the synaptic plasticity at a given time $t$ with a known $W(t_l)$ is:
    \begin{equation}
        W(t) = W(t_l)e^{\gamma(e^{t_j - t} - 1)}
    \end{equation}
    \begin{proof}
        Let $t_j$ be the post synaptic spike timing represented as a positive integer and let $t_l = t_j$. Assume all constants are positive real numbers. It is important to note that the bounds for synpatic effiacy $W$ are $0 < W < 1$.
        As there is no presynaptic activity, the equation for $dW(t)/dt$ becomes:
        \[
            \frac{dW(t)}{dt} = - \gamma e^{(t_j - t) / \Delta t} * W
        \]
        To solve for $W(t)$, the indefinite integral can be shown by the following:
        \begin{align*}
            \frac{dW(t)}{dt} &= -\gamma e^{(t_j - t) / \Delta t} * W \\
            \frac{dW(t)}{W} &= -\gamma e^{(t_j - t) / \Delta t} dt \\
            \int \frac{dW(t)}{W} &= \int -\gamma e^{(t_j - t) / \Delta t} dt \\
            -ln(|W(t)|) + C_1 &= -\gamma e^{(t_j - t) / \Delta t} + C_2\\
            ln(|W(t)|) &= \gamma e^{(t_j - t) / \Delta t} + C \quad \text{Note: } C = C_2 - C_1\\
            W(t) &= Ce^{\gamma e^{(t_j - t) / \Delta t}}
        \end{align*}
        To solve for the constant $C$, let $t = t_l$ as this is the synaptic weight values at the start of the synaptic update. Then, solving for $C$ entails the following:
        \begin{align*}
            W(t) &= Ce^{\gamma e^{(t_j - t) / \Delta t}} \\
            W(t_l) &= Ce^{\gamma e^{(t_j - t_l) / \Delta t}} \\
            W(t_l) &= Ce^{\gamma e^{(t_j - t_l) / \Delta t}} \\
            C &= W(t_l)e^{-\gamma} \\
            W(t) &= W(t_l)e^{-\gamma}e^{\gamma e^{(t_j - t) / \Delta t}} \\
            W(t) &= W(t_l)e^{\gamma(e^{(t_j - t) / \Delta t} - 1)} \\
        \end{align*}
    \end{proof}
\end{theorem}

Finally, as was done for the first TI-STDP theorem, we turn to considering the change in synaptic efficacy that results from TI-STDP in the scenario that only a post-synaptic spike occurs (but would still incur external stimulus beyond what comes from the one specific pre-synaptic neuron under consideration here). Formally, we only need to consider the negative synaptic shift induced by pre-synaptic time $t_i$ of zero and a non-negative post-synaptic $t_j$; this is proven as follows:
\begin{corollary}
    Given $0 = t_i < t_j$ then $W(t) \leq W(t_j)$. 
    \begin{proof}
        \begin{align*}
            W(t) &\leq W(t_j) \\
            W(t_l)e^{\gamma(e^{t_j - t} - 1)} &\leq {W(t_l)} \\
            e^{\gamma(e^{t_j - t} - 1)} &\leq 1 \\
            \gamma(e^{t_j - t} - 1) &\leq 0 \\
            e^{t_j - t} &\leq 1 \\
            t_j - t &\leq 0\\
            t_j &\leq t
        \end{align*}
    \end{proof}
\end{corollary}

%%% experiments and results %%%

\begin{figure}[!t]
    \centering
    \includegraphics[width=0.7\textwidth]{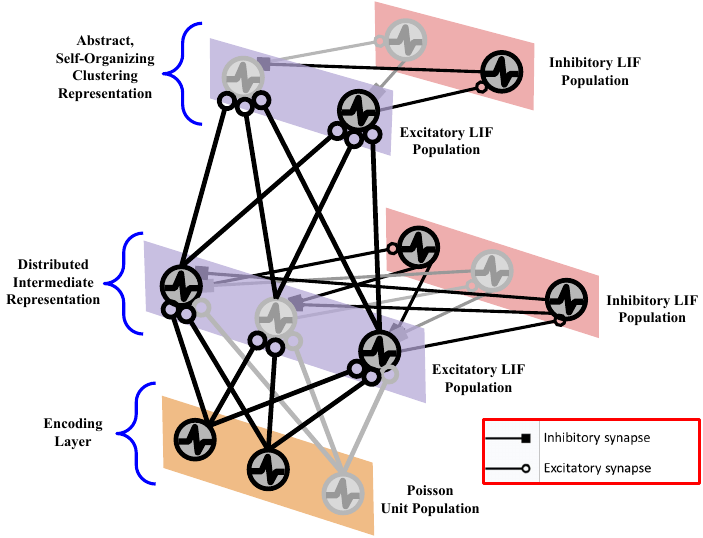} 
    \caption{Visual depiction of the biophysical spiking neural architecture that we simulate and analyze in our experimental study.} \label{fig:snn_arch} 
    \vspace{-0.4cm}
\end{figure}

\section{Experimental Results}
\label{sec:experiments}

\subsection{Predictive Generalization}

For this set of experiments, we examine predictive generalization ability of spiking neural models trained under different conditions of spike-timing centered plasticity. We utilize the MNIST database for the simulations carried for this part of the study. Specifically, this dataset consists of a large pool of gray-scale $28\times28$ images depicting hand-written digits across $10$ different categories (digits `0' through `9'). To train the spiking neuronal models, we construct a training set of $50,000$ samples and a validation (held-out) subset of $10,000$ that contains $1,000$ patterns from each class. The standard test-split of MNIST is used to evaluate the generalization performance of all models (with plasticity adjustment disabled). Note that the validation subset was used to manually tune/select hyperparameter values. The only pre-processing we used for these digits was to normalize their feature intensity values to lie in the range of $[0, 1]$ (by dividing image pixel values by $255$).

% \subsection{Clustering Model Setup}
For all generalization experiments, we use the same base model structure. Each model starts with a Poisson spike encoding sensory layer, which converts the real-valued image patterns on-the-fly to input spike trains -- these trains were driven by the normalized pixel intensities and constrained to never exceed a maximum spike frequency, i.e., never more than $64$ Hertz (Hz). Above the Poisson encoding layer are two hidden layers of recurrently-wired pairs of leaky integrate-and-fire (LIF) neurons, each of these acting as representation layers made up of a coupled set of excitatory and inhibitory units. The synapses connecting these the laterally-related neurons in any of these representation layers are fixed and configured such that the synaptic connections between the excitatory to the inhibitory units is set to a scaled identity matrix. Similarly, the connections from the inhibitory to the excitatory units was set to be a fixed, scaled hollow matrix. Finally, the synapses connecting the Poisson encoding input layer to the first representation layer, as well as the synapses connecting the first representation layer to the top-level second one, as plastic and adapted according to one of the forms of STDP we study in this work.\footnote{Note that the synapses carrying information across any two layers, e.g., sensory-to-hidden or hidden-to-hidden,  are specifically wired to the excitatory units of any representation layer.} In Figure \ref{fig:snn_arch}, we graphically depict the neuronal computational architecture simulated for this set of simulations. In Table \ref{tab:shared_params}, we describe, collect, and present the values of all essential, shared hyper-parameters that govern the models of this work for Case 1 and 2.

\paragraph{Experimental Cases} For this paper, we studied two different experimental settings/cases, each centered around different model learning conditions. %(in terms of the amount of data processed by a model). 
Note that the specific neuronal dynamics and configuration corresponding to each of these settings are discussed in the next section. In terms of specific learning conditions, we investigated the efficacy of models adapted under different forms of STDP, including our own proposed TI-STDP, when the amount of data seen was varied significantly. In the first experimental case, i.e., \textbf{Case 1}, each model was trained over $20$ passes (or epochs) over the MNIST \cite{lecun1998mnist} dataset, resulting in a data stream consisting of $250,000$ image patterns in total. In the second case, i.e., \textbf{Case 2}, models were trained strictly under one single epoch, resulting in a stream of only $50,000$ image patterns. The motivation for this particular condition is that it is essentially probing online learning performance, where each data point is only shown to the model once; such a setting which is far more neurocognitively realistic and representative of organism learning in the natural world.

\noindent
\textbf{Model Dynamics} As noted before, there were two different learning conditions that were examined in the experiments related to predictive generalization and each of these entailed the use of different neuronal dynamics; see Table \ref{tab:shared_params} for details. For Case 1, the neuronal dynamics of \cite{diehl2015unsupervised} were configured whereas, for Case 2, we configured a set of dynamics that were more neurobiologically realistic as well as challenging. For instance, in the first experimental case, the inhibitory neurons that feedback into the excitatory neurons use a scaling value of $-120$ which causes such a high-level of massive inhibition that it effectively turns any single layer of neurons into a near winner-take-all (sub-)systems. In contrast, the second experimental case uses a much smaller scale factor of $-10$, which promotes a less severe form of cross-layer competition and, as we observed in preliminary experimentation, induced more complex, temporally rich neuronal dynamics / spike trains.

\noindent
\textbf{Shared Model Parameters} For both experimental Case 1 and 2, there were certain parameter values that were shared across all of the different neuronal systems. Values for all the model components, excluding the synapses, were held constant trial-to-trial and learning rule-to-learning rule. There was one minor exception to this; empirically, we found that TI-STDP required a faster adaptive threshold time constant for Case 2 and we thus used this time constant value for simulations in both Case 1 and Case 2.

\begin{table}[!t]
    \centering
    \begin{tabular}{c|c|c|c} 
        \hline
        \textbf{Parameter} & \textbf{Case 1} & \textbf{Case 2} & \textbf{Description} \\ \hline 
        Input size & 784 & 784 & Dimensionality of sensory input \\ 
        \hline
        Hidden Layer Neurons & 625 & 625 & Dimensionality of $1$st representation layer\\ \hline
        Output size & 225 & 225 & Dimensionality of $2$nd representation layer\\ \hline
        Poisson (Maximal) Frequency & 63.75 & 63.75 & Maximum allowed input spiking frequency (Hertz)\\ \hline 
        $\tau_{me}$ & 100 & 100 & Excitatory membrane time constant (decivolts)\\ \hline
        $\tau_{mi}$ & 100 & 100 & Inhibitory membrane time constant (decivolts)\\ \hline
        $R_e$ & 100 & 100 & Excitatory resistance (deciOhms)\\ \hline
        $R_i$ & 100 & 100 & Inhibitory resistance (deciOhms)\\ \hline
        Refractory Time & 5 & 5 & Absolute refractory time (ms)\\ \hline
        $\tau^e_\theta$ & 1e5 & 1e5 & Excitatory threshold time constant (ms) \\ \hline
        $\tau^i_\theta$ & 0 & 0 & Inhibitory threshold time constant (ms) \\ \hline
        $\theta^+$ & 0.05 & 0.05 & Excitatory threshold increment  (decivolts)\\ \hline
        $\theta^e_0$ & -52 & -52 & Excitatory threshold base value  (decivolts)\\ \hline
        $\theta^i_0$ & -40 & -40 & Inhibitory threshold base value  (decivolts)\\ \hline
        $v^e_{rest}$ & -65 & -65 & Excitatory membrane resting potential  (decivolts)\\ \hline
        $v^i_{rest}$ & -60 & -60 & Inhibitory membrane resting potential  (decivolts)\\ \hline
        $v^e_{reset}$ & -60 & -60 & Excitatory membrane reset potential  (decivolts)\\ \hline
        $v^i_{reset}$ & -45 & -45 & Inhibitory membrane reset potential  (decivolts)\\ \hline
        $W_{ei}$ & 22.5 & 22.5 & Excitatory-to-inhibitory synaptic scale\\ \hline
        $W_{ie}$ & -120 & -10 & Inhibitory-to-excitatory synaptic scale\\ \hline
        $R_1$ & 1 & 1 & Input-to-hidden resistance (deciOhms)\\ \hline
        $R_2$ & 6 & 6 & Hidden-to-hidden resistance (deciOhms)\\ \hline
    \end{tabular}
    \vspace{0.15cm}
    \caption{\textbf{Shared Model-Level Hyper-parameters.} Here we describe and provide chosen values for the key meta-parameters that govern the models simulated for the predictive generalization experiments that were conducted with the MNIST database.}
    \label{tab:shared_params}
    \vspace{-0.4cm}
\end{table}

\begin{table}[!t]
    \centering
    \begin{tabular}{c|c|c||c|c|c||c|c|c}
      \hline
      \multicolumn{3}{c||}{\textbf{TR-STDP}} & \multicolumn{3}{c||}{\textbf{EV-STDP}} & \multicolumn{3}{c}{\textbf{TI-STDP}} \\
      \textbf{Parameter} & \textbf{Case 1} & \textbf{Case 2} & \textbf{Parameter} & \textbf{Case 1} & \textbf{Case 2} & \textbf{Parameter} & \textbf{Case 1} & \textbf{Case 2}\\ \hline 
      $\tau_{z}$ & 20 & 20 & $\eta_w$ & 1 & 0.01 & $\alpha$  ($\ell=1$) & 0.00375 & 0.00375 \\ \hline
      $z^{\ell=1}_{tar}$ & 0.3 & 0.3 & $A^+$ ($\ell=1$) & 0.0055 & 1 & $\alpha$ ($\ell=2$) & 0.05 & 0.025 \\ \hline
      $z^{\ell=2}_{tar}$ & 0.025 & 0.025 & $A^-$ ($\ell=1$) & 0.001375 & 0.3 & $\hat{\beta}$ ($\ell=1$) & 1.25 & 1.25 \\ \hline
      $A^+$ & 0.01 & 0.01 & $A^+$ ($\ell=2$) & 0.0055 & 1 & $\hat{\beta}$  ($\ell=2$) & 2 & 2 \\ \hline
      $A^-$ & 0.001 & 0.001 & $A^-$ ($\ell=2$) & 0.000275 & 0.075 & $\hat{\gamma}$ ($\ell=1$) & 0.75 & 0.75 \\ \hline
        &  &  & $\lambda$ & 0 & 0 & $\hat{\gamma}$  ($\ell=2$) & 0.125 & 0.25 \\ \hline
    \end{tabular}
    \vspace{0.1cm}
    \caption{Hyper-parameter value configurations for each plasticity model studied in this work. Note that we have further tagged certain meta-parameter with layer $\ell$ indicators for cases where values were different for different neuronal layers (otherwise, the value reported was set to be the same across all layers).}
    \label{tab:hyperparam_config}
    \vspace{-0.4cm}
\end{table}

\noindent 
\textbf{Plasticity-Specific Parameters} When implementing TI-STDP into an artificial neuronal system some specific algorithm design choice swere made. None of these changes affect the dynamics shown in section \ref{sec:TISTDP} but were carefully chosen so as to make tuning a TI-STDP-adapted system easier. A key change we found important was the introduction of the additional coefficient $\alpha$, which acted as a tunable global learning rate. In relation to the proposed plasticity equations \ref{eqn:quietTI} \ref{eqn:noisyTI}, the coefficients $\gamma$ and $\beta$ can then be reformulated as: $\gamma = \alpha \hat{\gamma}$ and $\beta = \alpha \hat{\beta}$. This reformulation was done to provide a simple control factor that controlled fast a model's synapses adapt without having to adjust $\gamma$ and $\beta$ individually (as manipulating these required greater care).

Also while tuning neuronal systems trained with TI-STDP, a key experimental insight we found was to pay due attention to the ratio between the hyper-parameters $\beta$ and $\gamma$. Empirically, we found that this ratio, when set correctly, causes each synapse to reach a steady state based on the frequency of pre- and post-synaptic activity. In general, if more post-synaptic activity is desirable, then increasing this ratio will cause the steady state to be achieved at a higher degree of synaptic efficacy/strength. 
In turn, if less post-synaptic activity is desirable, then decreasing this ratio will work well to achieve the opposite effect. The abstraction of $\alpha$ notably allows for the experimentalist/modeler to directly control the speed of learning to be changed while keeping this ratio constant.

%% TODO: consider replacing cluster metrics w/ supervised classifier probe instead?
\begin{table}[!t]
\begin{center}
\begin{tabular}{c | c c c } 
 \hline
 \textbf{MNIST} & \multicolumn{3}{c}{\textbf{Case 1: Multi-Pass Dynamics}} \\
 %\textbf{MNIST} & \multicolumn{3}{c}{Supervised Measurements} \\
 \textbf{Model} & \textbf{Acc} & \textbf{Precision} & \textbf{Recall} \\ %[0.5ex] 
 \hline\hline
 TR-STDP & $78.190 \pm 0.320$ & $78.301 \pm 0.137$ & $77.878 \pm 0.376$ \\ %[1ex] 
 EV-STDP & $84.983 \pm 0.937$ & $85.172 \pm 0.661$ & $84.783	\pm 0.912$ \\ %[1ex] 
 TI-STDP & $82.093 \pm 0.478$ & $82.003 \pm 0.514$ & $81.785 \pm 0.485$ \\ %[1ex] 
 \hline
 & \multicolumn{3}{c}{\textbf{Case 2: Online  Dynamics}} \\
 \hline\hline
 TR-STDP & $68.973 \pm 0.626$ & $70.777 \pm 0.631$ & $68.110 \pm 0.652$ \\ %[1ex] 
 EV-STDP & $62.800 \pm 2.418$ & $68.717 \pm 4.619$ & $61.945 \pm 2.543$ \\ %[1ex] 
 TI-STDP & $70.123 \pm 1.131$ & $70.309 \pm 0.852$ & $69.450 \pm 1.157$ \\ %[1ex] 
 \hline
\end{tabular}
\end{center}
\caption{\textbf{Model Generalization performance measurements.} Reported are the mean and standard deviation for three supervised and three unsuperivsed evaluation metrics over three simulation trials of the biophysical network studied in this work. Supervised predictions are obtained from the unsupervised biophysical models by binding the category labels to specific neurons during a post-training phase and allowing each neuron in the top-layer vote for the class it believes a sensory input pattern belongs to.}
\label{tab:generalization_results}
\vspace{-0.4cm}
\end{table}

\begin{figure}[!t]
    \centering
    \begin{subfigure}[b]{0.475\textwidth}
        \includegraphics[width=1\textwidth]{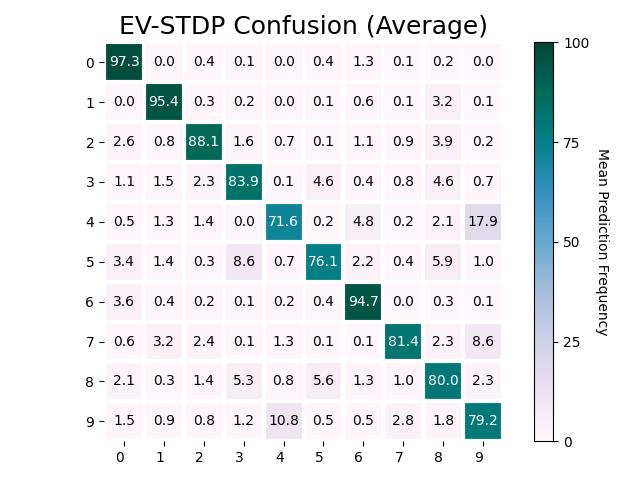} 
        %caption{EV-STDP.}
    \end{subfigure}
    \begin{subfigure}[b]{0.475\textwidth}
        \includegraphics[width=1\textwidth]{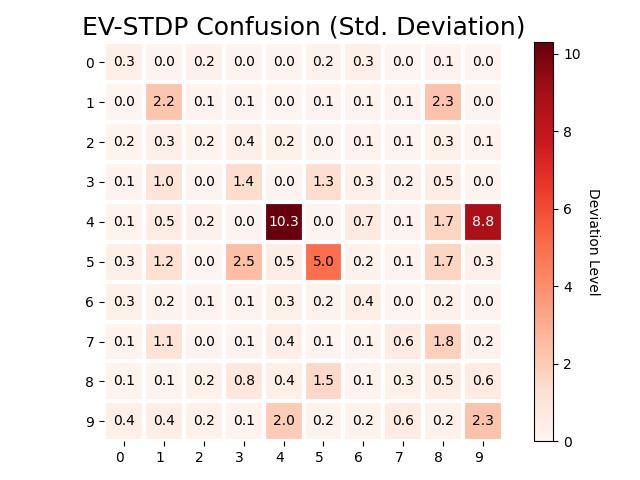}
        %\caption{EV-STDP Confusion.}
    \end{subfigure}\\
    \begin{subfigure}[b]{0.475\textwidth}
        \includegraphics[width=1\textwidth]{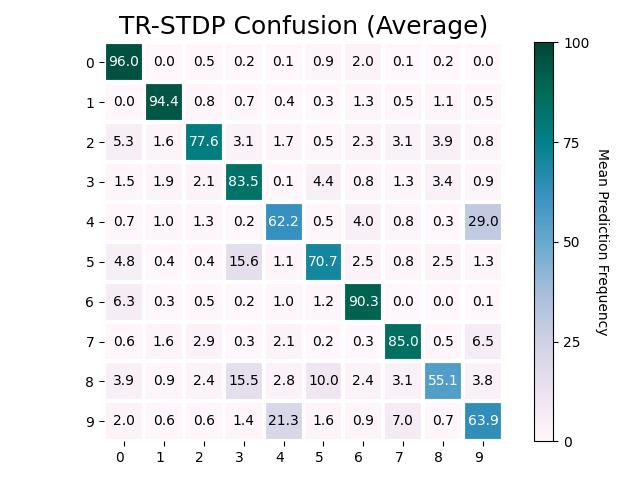}
        %\caption{TR-STDP Confusion.}
    \end{subfigure}
    \begin{subfigure}[b]{0.475\textwidth}
        \includegraphics[width=1\textwidth]{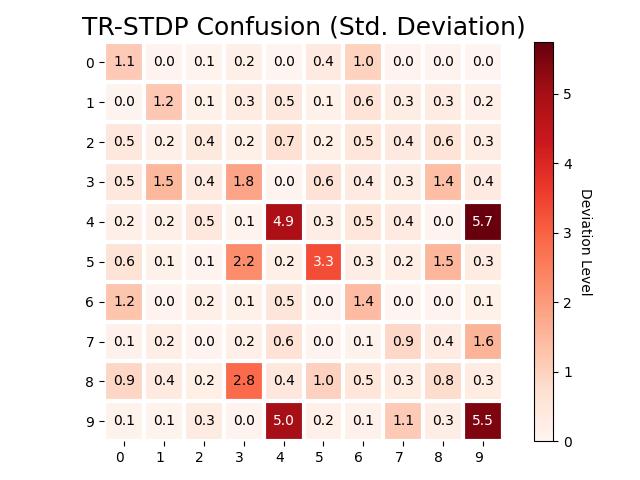}
        %\caption{TR-STDP Confusion.}
    \end{subfigure}\\
    \begin{subfigure}[b]{0.475\textwidth}
        \centering
        \includegraphics[width=1\textwidth]{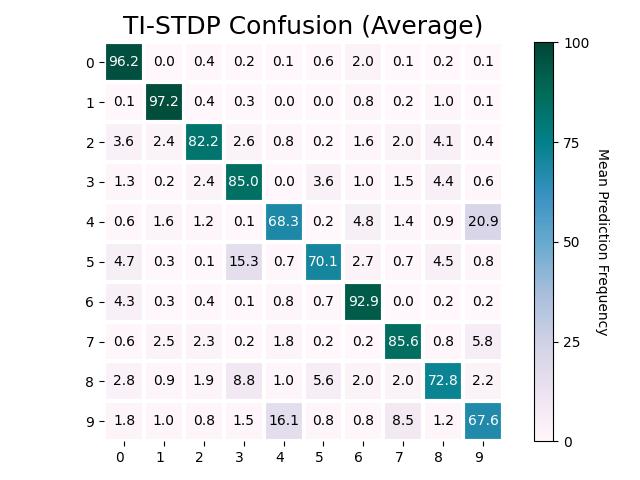}
        %\caption{TI-STDP Confusion.}
    \end{subfigure}
    \begin{subfigure}[b]{0.475\textwidth}
        \centering
        \includegraphics[width=1\textwidth]{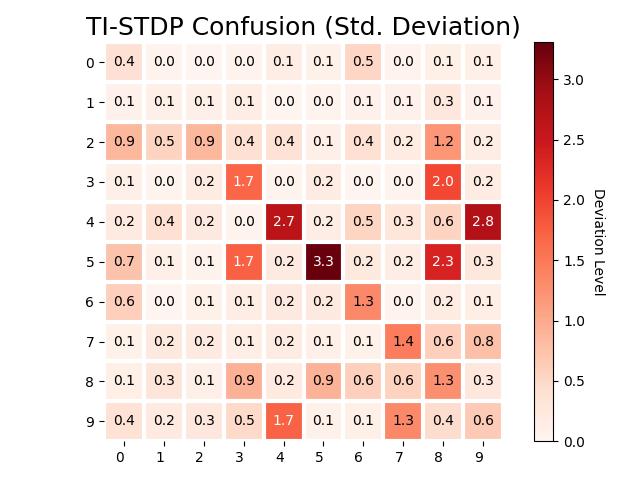}
        %\caption{TISTDP Confusion.}
    \end{subfigure}
    \caption{\textbf{Confusion heat-map visualization.} Shown are the heat-maps for the mean (`Average') and standard deviation (`Std. Deviation') of the confusion matrices for each of the three plasticity models; plot order arranged from top to bottom - event-based post-synaptic STDP (EV-STDP), trace-based STDP (TR-STDP), and time-integrated STDP (TI-STDP).} 
    \label{fig:confusion}
    \vspace{-0.4cm}
\end{figure}

\subsection{Clustering, Binding, and Performance Measurements} 
\label{sec:binding_measurements}

While these models are unsupervised in nature, working to cluster the patterns they process in their upper-level latent space, we may equip them with a simple mechanism for conducting classification so that we may probe their generalization ability. Similar to the scheme mentioned and utilized in \cite{diehl2015unsupervised}, after adapting each neuronal model under a particular plasticity scheme, we `bind' a subset of the labels available in the database used for training to the LIFs in the top-most layer. Specifically, to bind labels to neuronal units and conduct classification, we adhere to the following process:
\begin{enumerate}[noitemsep,nolistsep]
    \item For each presented sensory input pattern, the firing response of each LIF in the top-most layer, which contains $H$ neurons, is recorded and its corresponding spike frequency over the $T$-length stimulus window is stored in relation to the label $y \in \{0,1,2,..,Y\}$ of that particular sensory input. Formally, this means for any neuron $k$ in the second layer $\ell = 2$ of our spiking model, we record the spike frequency $\lambda_k = \sum^T_{t=1} s^{\ell=2}_k$ induced by an observed input-label pair $(\mathbf{o}, y)$, and add it to a global frequency matrix $\mathbf{C}_{y,k} \in \mathbb{Z}_{\geq 0}^{Y \times H}$ where the class label $y$ indexes the correct row of the matrix and $k$ indexes the appropriate neuronal unit.
    \item After every data point in the sub-sample is processed by the spiking model, and we have aggregated all responses, we then take the $\arg\max$ across all $Y$ rows to obtain the class that each neuron $k$ responds the most to and assign or `bind' this winning class index to neuron $k$.
    \item During test-time inference, with long-term plasticity (and adaptive thresholds) disabled, we run an unseen input pattern $\mathbf{o}$ through the model and record the spike frequency vector $\vec{\mathbf{\lambda}}$ for all neurons in layer $\ell = 2$. To obtain a predicted class index, we do the following: $\arg\max_{i \in \{1,2,...,H\}} \mathbf{\lambda}_i | \lambda_i \in \vec{\mathbf{\lambda}}$, which returns the index of the neuron with the highest response to the input pattern from which we may extract the bound class index from the previous step.
\end{enumerate}
In the experiments of this paper, for all neuronal models, we bind labels to neurons in the second layer using the last $10000$ training samples encountered during model training. 

In Table \ref{tab:generalization_results}, we report the generalization ability of each model, when using the classification scheme presented above. We specifically measure the model's predictive accuracy, precision, and recall on the $10000$ held-out test-set image samples; all values reported in Table \ref{tab:generalization_results} are the mean and standard deviation over three experimental simulation trials (where each simulation run was seeded with a unique integer). Finally, we observe the types of mistakes made by the various simulated models by recording their test-set normalized confusion matrices. Concretely, we visualize the mean and standard deviation normalized confusion matrices (across the three trials) for each model as heat-maps in Figure \ref{fig:confusion}; for the mean confusion heat-map, a stronger, darker green indicates better performance (closer to $100$\% prediction performance on a specific class category) whereas for the standard deviation confusion heat-map, a stronger, darker red indicates more statistical variability in the model's predictive outputs.

\begin{figure}[!t]
    \centering
    \begin{subfigure}[b]{0.32955\textwidth}
        \includegraphics[width=1\textwidth]{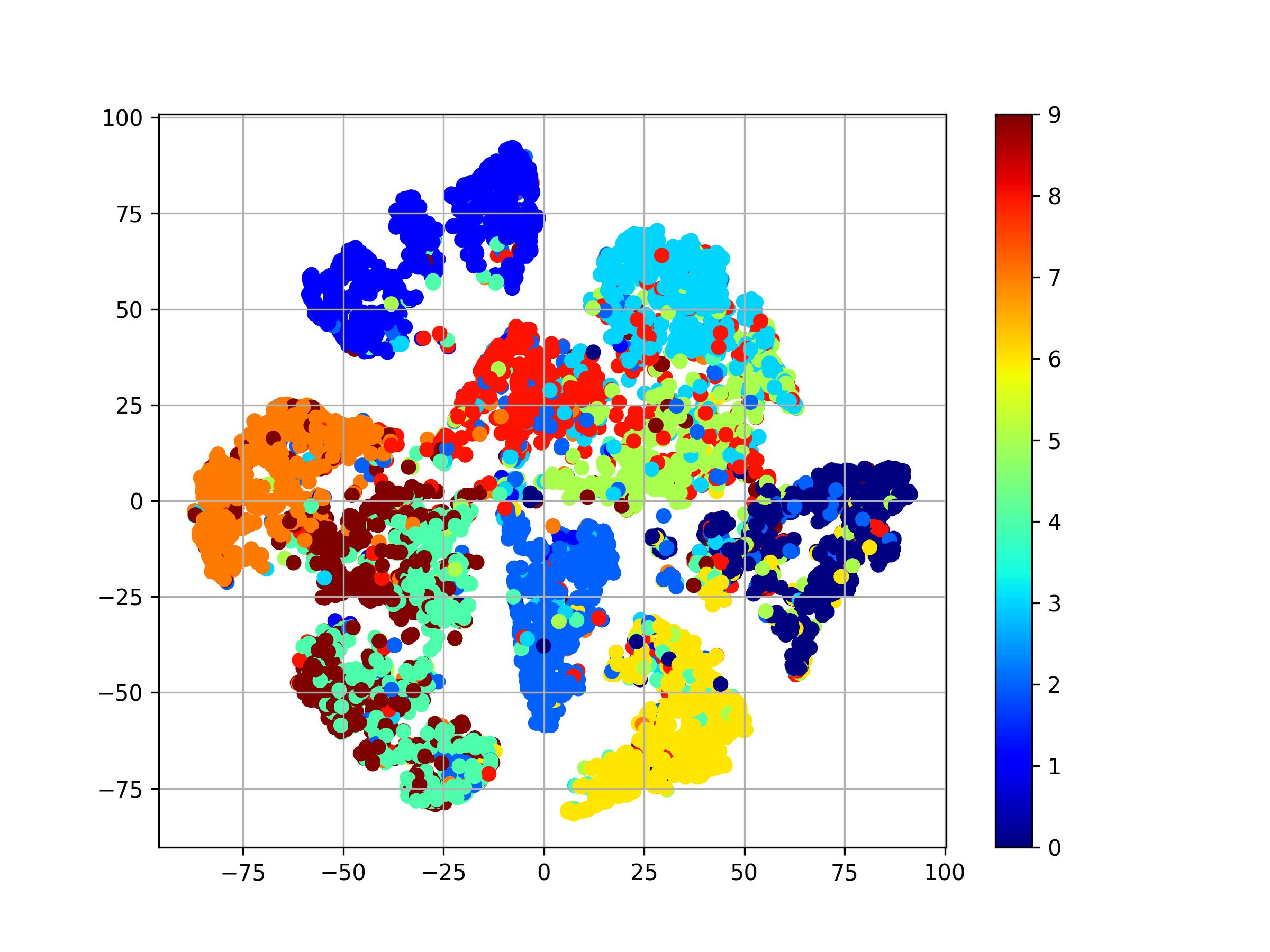} 
        \caption{TR-STDP Latent Clusters.}
    \end{subfigure}
    \begin{subfigure}[b]{0.32955\textwidth}
        \includegraphics[width=1\textwidth]{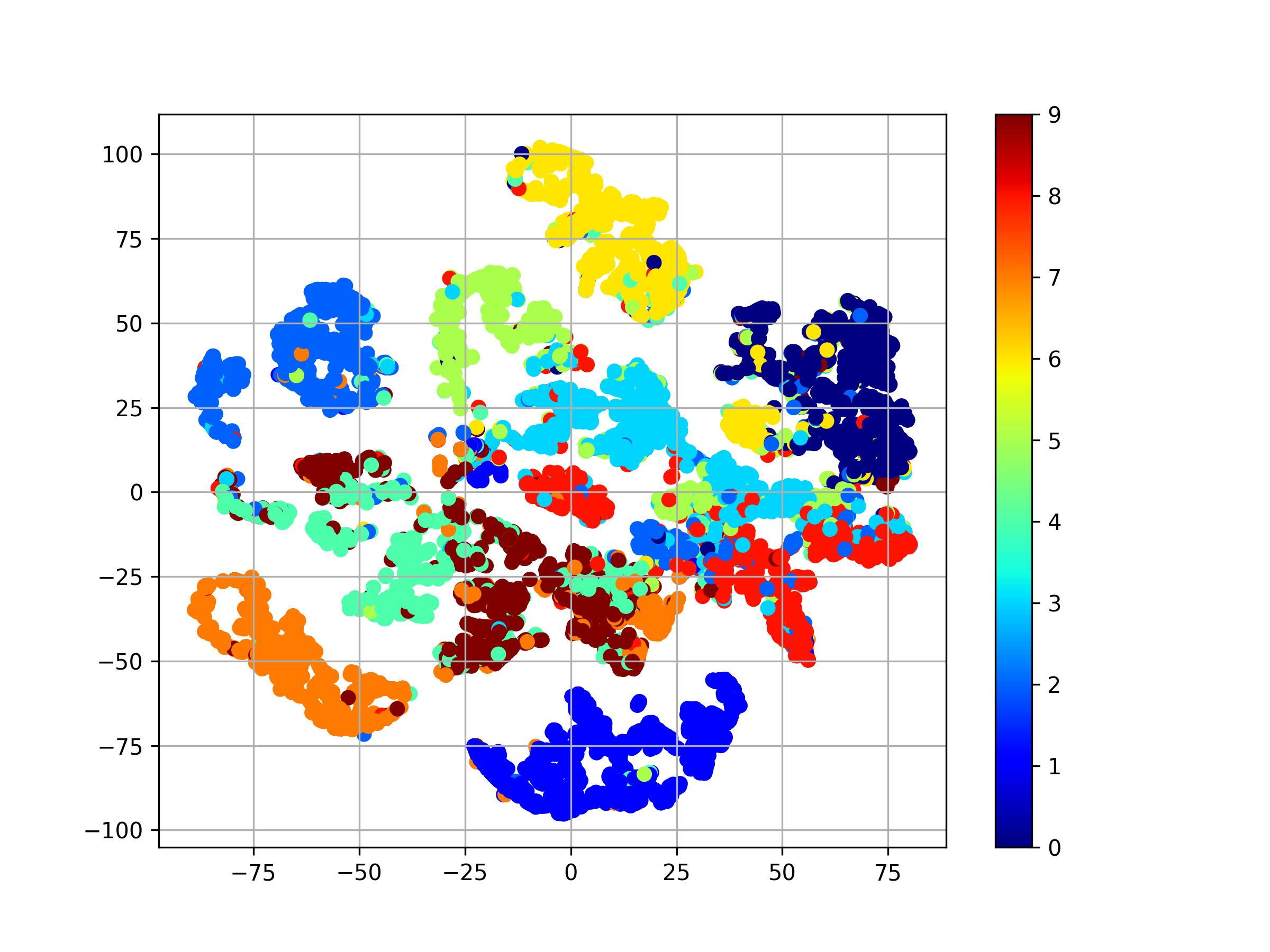}
        \caption{EV-STDP Latent Clusters.}
    \end{subfigure}
    \begin{subfigure}[b]{0.32955\textwidth}
        \includegraphics[width=1\textwidth]{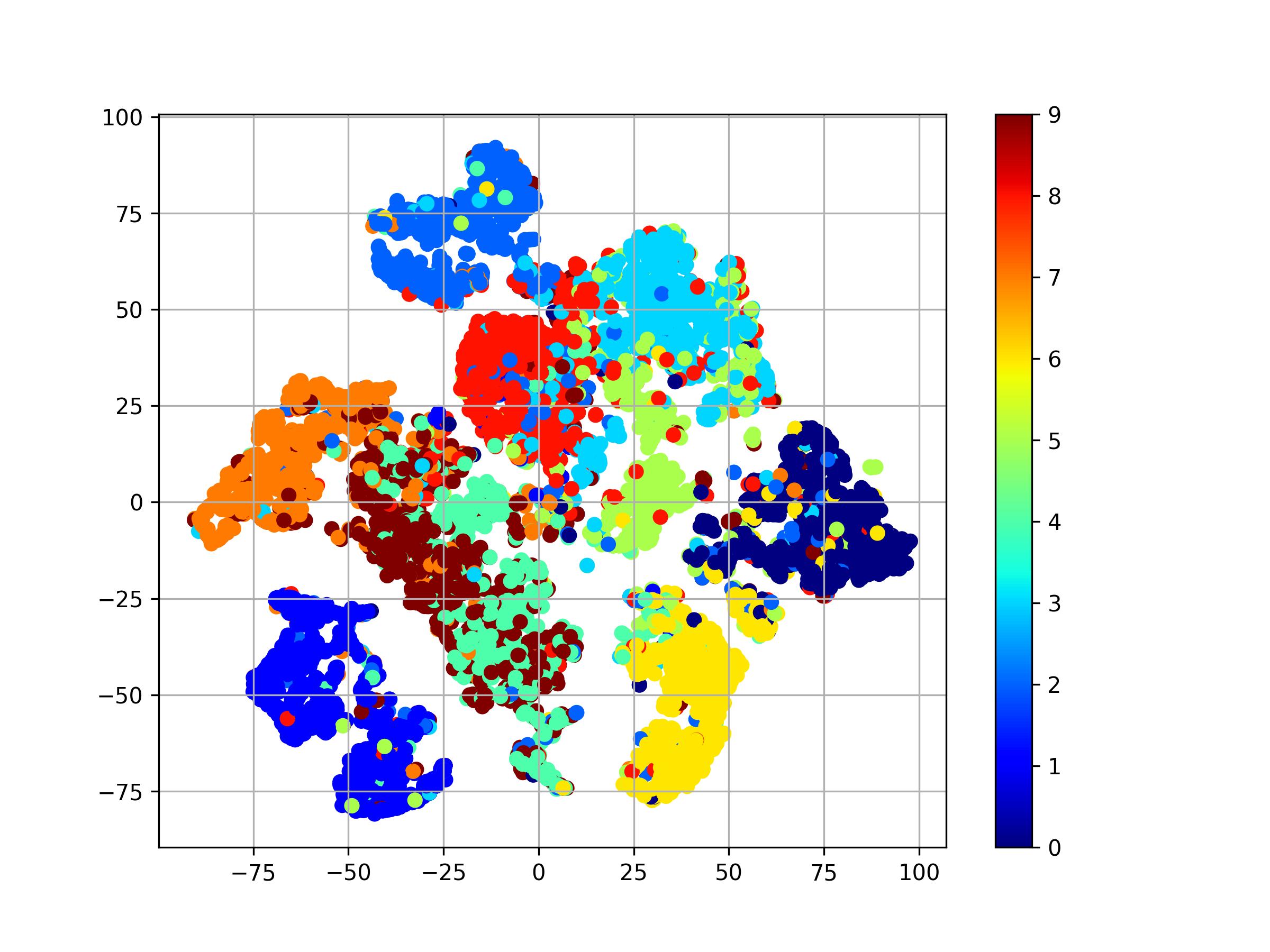}
        \caption{TI-STDP Latent Clusters.}
    \end{subfigure}
    \caption{\textbf{Latent t-SNE Plots.} t-SNE Visualization of sampled clusters that were extracted under different spike-timing-driven algorithm conditions.} 
    \label{fig:tsne}
    \vspace{-0.5cm}
\end{figure}

Finally, in Figure \ref{fig:tsne}, we examine the latent clusters formed by spiking model's topmost layer, under each spike-centered plasticity condition. To visualize these implicit clusters, we collected the neural latent `codes' that characterize the topmost layer by converting the temporal spike train produced in response to each presented pattern to an approximate rate-code vector $\mathbf{c}^{\ell=2}$ in the following manner:
\begin{equation}
    \mathbf{c}^{\ell=2} = (\gamma_c/T) \sum^T_t \mathbf{s}^{\ell=2}(t) 
\end{equation}
where we set $\gamma_c = 1$. After feeding in all of the data patterns contained in the dataset's test-set and collecting all of the resultant rate codes, we visualized the emergent rate-code islands using the t-Distributed Stochastic Neighbor Embedding (t-SNE) algorithm \cite{van2008visualizing}. Qualitatively, we see that all three forms of STDP form useful clusters that represent the different categories inherent to the MNIST database, with EV-STDP forming slightly better clusters (less overlap, as opposed to TR-STDP and TI-STDP) which explains its slightly better performance than TI-STDP. Note that, in the supplementary material, we provide an additional visualization that further probes the clusters that form as a result of the top-most layer's self-organization.

\subsection{Qualitative Patch Extraction Results}
\label{sec:patch_model}

%% to merge with below
An important aspect of human visual perception vision is its ability objects within scenes and effectively representing them in terms of structures \cite{robertson1991neuropsychological} often referred to as `part-whole hierarchies' \cite{hinton1979some}. The foundational neurocognitive premise is that an `object' can be regarded as a collection or combination of parts (and those parts are made up of sub-parts \cite{hinton1979some}, etc.), with particular part-to-part relationships (which constitutes their overall organization). In particular, perception involves decomposing the observed nested structure in one's niche into its base parts or components; this results in a form of compositionality \cite{lake2015human,fisher2022recursive}, where the same building block patterns can be recursively and hierarchically arranged to assemble a vast array of possible complex patterns. Building intelligent agents that are capable of learning part-whole hierarchical representations facilitates a wide variety of cognitive functionality, including planning or confabulating novel configurations of parts (and sub-parts) as well as recognizing previously unseen configurations (in the context of zero-shot generalization). 

\begin{figure}[!t]
    \centering
    \includegraphics[width=0.8\textwidth]{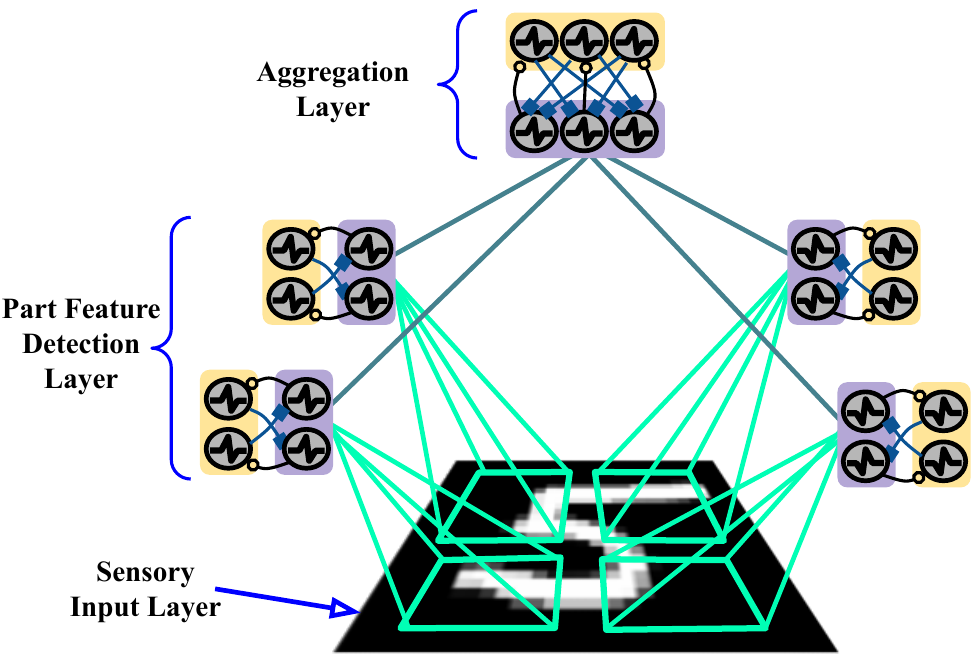} 
    \caption{Visual depiction of the biophysical spiking compositional model that we designed and studied.} 
    \label{fig:compositional_model} 
    \vspace{-0.4cm}
\end{figure}

Motivated by the above, we next constructed neuronal models that focused on representation learning, utilizing our proposed TI-STDP for synaptic adaptation. The biophysical dynamical system we designed was meant to learn how to represent the sensory inputs it encounters in a ``constructive fashion''. Specifically, this was enforced by encoding a `local connectivity' inductive bias into the first layer's synaptic connectivity pattern, i.e., the first set of synaptic connections of our model employed a partitioned matrix design such that groups/clusters of neuronal units -- specifically coupled excitatory-inhibitory neurons -- in the first layer would only process particular patches of a sensory input at particular spatial locations (with some small possible degree of overlap). 
The spike train activities emitted by these local groups of coupled neurons would then be aggregated by a second layer of coupled LIF neurons. The intuition behind this three-layer model (sensory input, layer of local neuronal spiking feature detectors, and an aggregation neuronal layer) was that the first would specialize in extracting low-level features, such as strokes or pattern arcs/curves, while the second layer would learn to ``combine'' these lower-level features into higher-level primitive objects (e.g., digits from strokes/edges). The synapses connecting the sensory layer -- a set of spatially arranged $7 \times 7$ pixel image patches -- to the intermediate layer as well as those connecting the intermediate layer groups to the aggregation layer were shaped by the plasticity dynamics of long-term potentiation, depression, and synaptic decay induced by TI-STDP.

See Figure \ref{fig:compositional_model} for a visual depiction of our biophysical, compositional model. In terms of parameterization, our model's sensory input layer entailed partitioning an input sensory image into a set of $7\times7$ patches (with a configurable pixel overlap as high as $2$) that were each encoded dynamically into Poisson spike trains constrained to have a maximum frequency of $63.75$ Hz. The intermediate layer consisted of $16$ groups of $64$ excitatory units coupled to inhibitory units (two sets of $1024$ neurons total) while the aggregation layer consisted of $225$ excitatory units coupled with $225$ inhibitory units.

To test if the above three-layer compositional model learned some aspect of part-whole relationships from the data, after training it on $250,000$ images (or five passes through the MNIST database), we allowed the model to build/compose patterns by constructing a simple top-down procedure that we call \emph{ancestral assembly}. This pattern construction process adhered to the following two steps:
\begin{enumerate}[noitemsep,nolistsep]
    \item \textbf{Part Feature Selection:} For any given LIF neuron $k$ in the aggregation (second) layer $\ell=2$, the top $K = 300$ synapses with the highest efficacies from $\mathbf{W}^{\ell=2}$ were selected, which when backwards traversed, returned the indices of the neurons in the intermediate representation layer that correlated most strongly with the aggregation neuron $k$, i.e., the set $\mathcal{Q}$; 
    \item \textbf{Object Assembly:} The receptive field of each neuron $q \in \mathcal{Q}$ was extracted, modulated by the synaptic efficacy connecting it to aggregation neuron $k$, and added to a scratch-pad image (initialized to a $D \times D$ matrix of zeros). Formally, an object pattern $\mathbf{\tilde{o}}$ was assembled via the following weighted super-position: 
    $\mathbf{\tilde{o}} = \sum_{q \in \mathcal{Q}} W^{\ell=2}_{k,q} \mathbf{W}^{\ell=1}_{q,:}$, where $\mathbf{W}^{\ell=1}_{q,:}$ denotes the `slicing out' of the $q$th row of synaptic matrix $\mathbf{W}^{\ell=1}$.
\end{enumerate}
In effect, the above two-step scheme represents an approximate way utilize the spiking neuronal circuit's underlying generative process, without conditioning it on any context information, to produce sensory patterns that provide some indication/interpretation of the knowledge it has acquired. 

\begin{figure}[!t]
    \centering
    \begin{subfigure}[b]{0.45\textwidth}
        \includegraphics[width=1\textwidth]{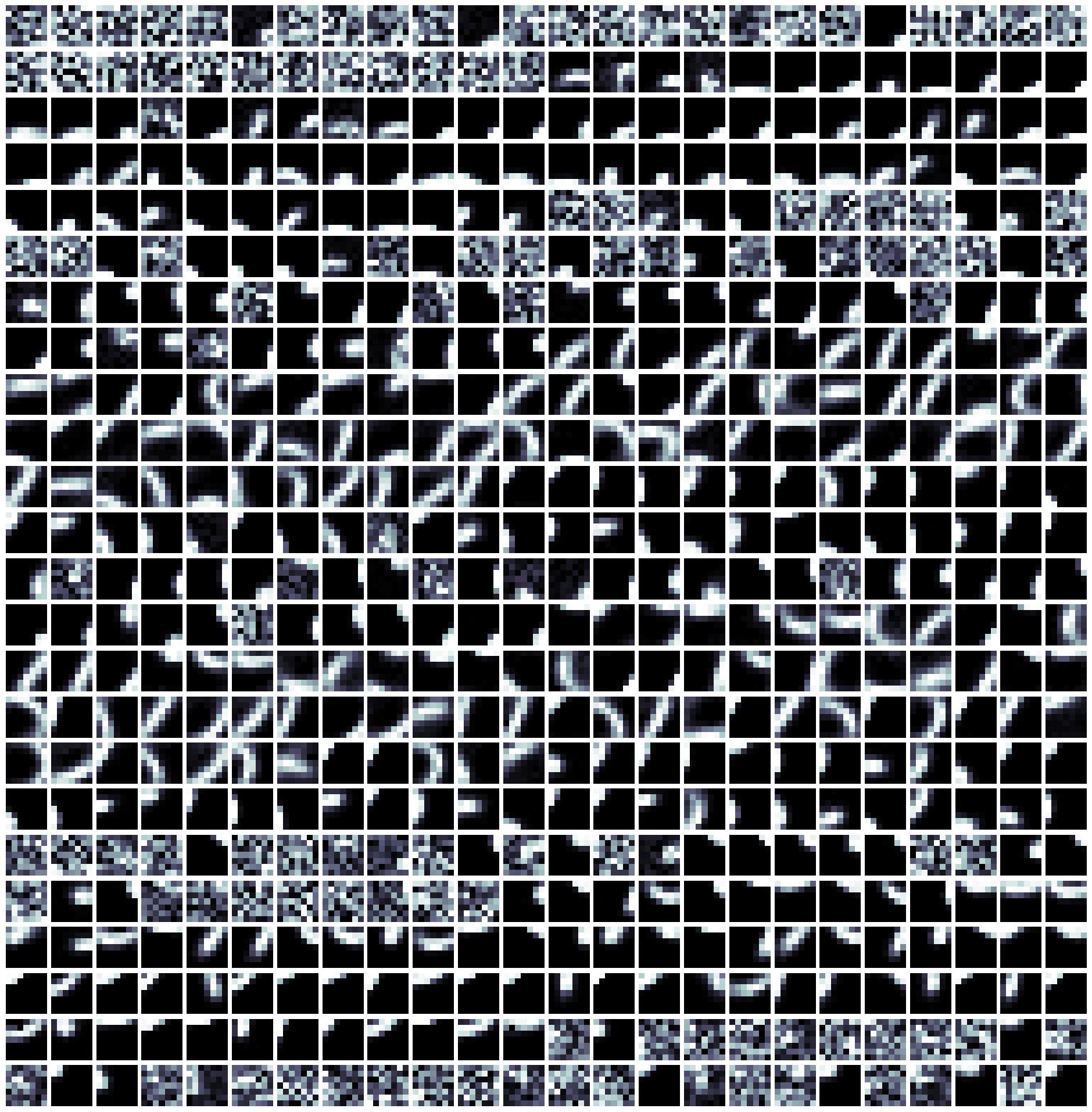} 
        \caption{Sample of Sensory-Level Filters.}
        \label{fig:receptive_fields}
    \end{subfigure}
    \begin{subfigure}[b]{0.45\textwidth}
        \includegraphics[width=1\textwidth]{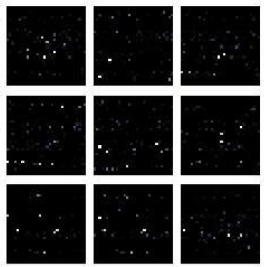} 
        \caption{Sample of Aggregation-Level Synapses.}
        \label{fig:aggregation_fields}
    \end{subfigure}\\
    \begin{subfigure}[b]{0.45\textwidth}
        \includegraphics[width=1\textwidth]{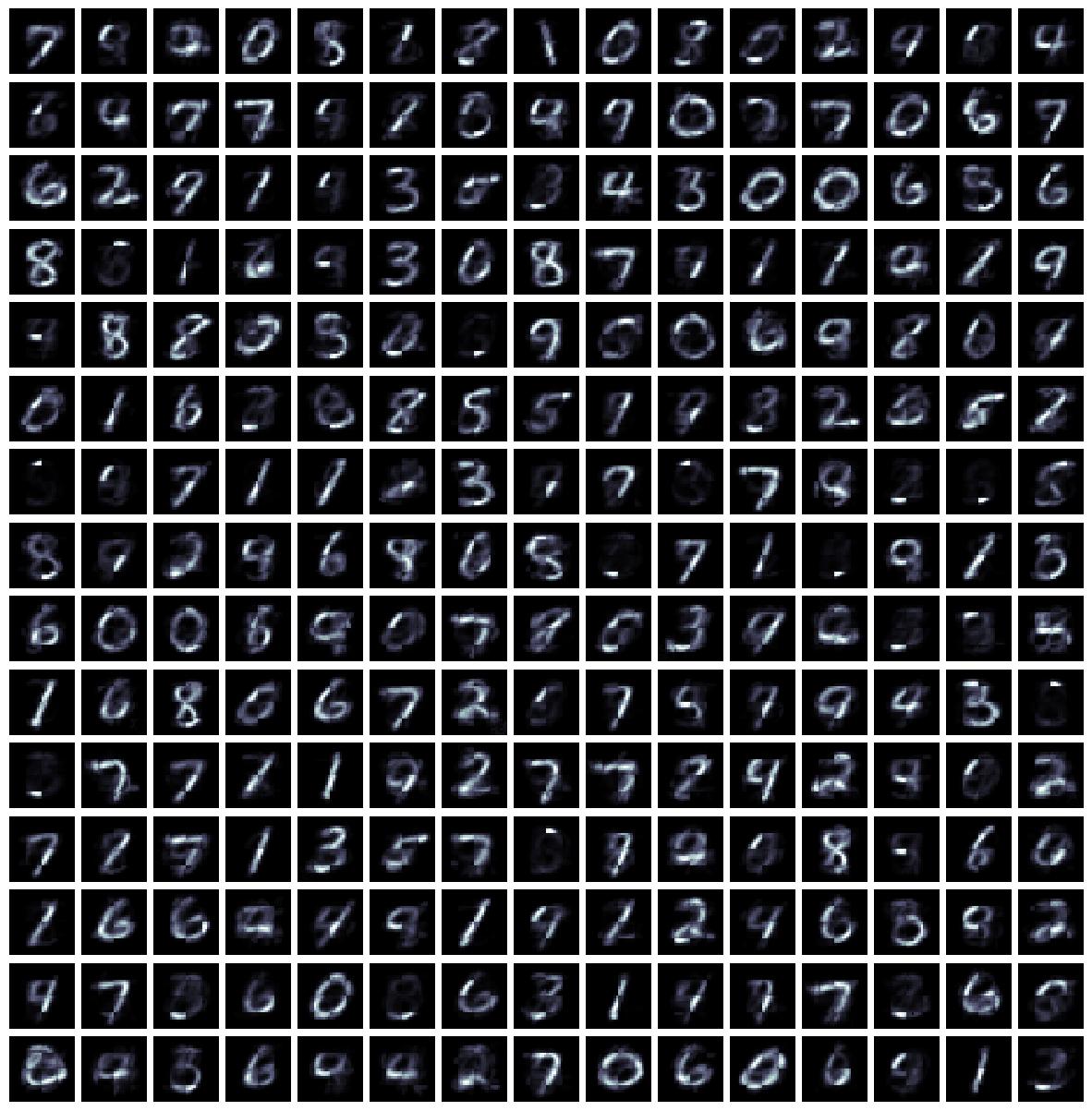}
        \caption{Digit Object Assembly Samples.}
        \label{fig:assembled_objects}
    \end{subfigure}
    \caption{\textbf{Visualization of the spiking compositional model's learned elements.} For our TI-STDP adapted compositional, hierarchical biophysical model, we visualize the following: sub-Figure \ref{fig:receptive_fields} contains the model's acquired sensory-level receptive fields (which contain strokes/arcs of different rotations/shifts),  
    sub-Figure \ref{fig:aggregation_fields} contains the model's top-level synapses (brighter white values indicate stronger synaptic efficacies) and, 
    sub-Figure \ref{fig:assembled_objects} shows objects that have been ``crafted'' by the learned model using our ancestral assembly process.} \label{fig:compositional_results}
    \vspace{-0.4cm}
\end{figure}

We visualize the sampled assembled patterns produced by our compositional model in Figure \ref{fig:assembled_objects}. Notice that, desirably, the compositional model was able to synthesize or `confabulate' at least one of every possible digit type it processed in the MNIST database, demonstrating its ability to construct complete digit patterns from the lower-level stroke arcs its intermediate neuronal feature detector groups acquired. In Figure \ref{fig:receptive_fields}, we present a large subset of the receptive fields the neuronal feature detectors acquired in the first layer; in Figure \ref{fig:aggregation_fields}, we present a subset of synaptic connectivity patterns acquired in the second layer. Notice that the second layer ends up being shaped by TI-STDP to be sparse, demonstrating that a basic latent command structure emerges where the second layer of coupled LIFs learn to control and compose the receptive fields acquired by the groups of intermediate coupled LIFs.

% In this section, you generally start by discussing the environment setup (task, hardware, software, etc.) as well as the details related to the datasets you are using or task environments that you are simulating. Fully specify the comparison setup -- do not forget to define formally/mathematically the metrics you will use, the mechanics and key notions that characterize the different methods/baselines you will be comparing to, as well as the details behind the ablation of different components if you have a modular, multi-component system or model. If you are using any well-defined public tasks and datasets, it is expected that you still define the details of each in your own words (however, when space is limited for the paper, as in a conference, you can offload this to an appendix section). 

% Do not forget to include a small sub-section on your hardware setup (either the properties of your desktop or the servers/computing clusters that you make use of) if you are doing computational work.

%Benchmarks
%Experimental Setup
%Results

\section{Discussion}
\label{sec:discussion}

\paragraph{Discussion:} The clustering experiments done in this paper were designed to examine and compare the effectiveness of different spike-timing-dependant plasticity-based synaptic adjustment methods. The first experiment, i.e., Case 1, was carried out to demonstrate that the spiking neural circuits do not collapse and achieve stable performance when adapting with longer streams of data patterns (as emulated by multiple epochs over the database). The second experiment, i.e., Case 2, focused on adaptation that is carried out across a single epoch of the data, which emulated aspects of online learning, motivated by the fact that online learning better embodies natural, real-world learning than credit assignment carried out over multiple epochs.

There were three different synaptic update methods that were studied in this paper: trace-based STDP, event-based STDP, and time-integrated STDP. All of them were hand-tuned to attain the performance measurements we reported across both experimental cases. Overall, we found experimentally that the predictive generalization across synaptic plasticity algorithms was quite close with EV-STDP performing the best (with TI-STDP coming in a close a second place) when run with many passes over the data whereas TI-STDP performed the best in terms of online learning.

Looking to the t-SNE plots of each of the trained models \ref{fig:tsne} in tandem with their confusion matrices \ref{fig:confusion}, we observe that each type of synaptic plasticity model does a fairly good job at separating out the different digits. From the t-SNE plots, it is clear that some digit patterns, such as those that classify as fours and nines, prove to be difficult for the adapted spiking model to learn to separate, which intuitively makes sense given that many of the fours in MNIST exhibit a large degree of overlap with the nines. This difficulty due to pattern overlap can also be observed, to a lesser extent, with respect to the threes, fives, and eights. Note that these label-based separation difficulties are further reflected in the kinds of categorization mistakes each model makes, where, in the confusion matrices presented in Figure \ref{fig:confusion}, we observe that there are as many predictions of fours as nines and nines as fours.

\noindent 
\textbf{Limitations} Note that, while our empirical results demonstrate promise, there are several limitations in this work worth pointing out. First and foremost, there are known limitations to constructing effective SNNs, specifically related to the fact that they are slow to train and often difficult to tune. We note that each experiment in this paper took well over a day per model to run across all the trials (using a single performant GPU). TI-STDP does not work to reduce this cost as the linear algebra needed to simulate its underlying mechanics is not any less complex than the other studied methods. It could prove fruitful to consider simplifications of TI-STDP that reduce the number of required operations to perform useful synaptic adjustments (such as the binary STDP rule developed in \cite{ferre2018unsupervised}), which would prove useful for instantiation in in-memory hardware platforms, e.g., neuromorphic chips. 

In addition to the problem of speed mentioned above, all models used here require tuning to get them to function properly and stably. Both TR-STDP and TI-STDP were found, experimentally in this work, to be significantly less fragile when it comes to hyper-parameters; however, we remark that the phrases ``not fragile'' as opposed to ``robust'' are far from the same thing. Another important experimental, practical limitation we observed was that, if weight-dependency scaling was not used in any plasticity-case, it was more likely that the spiking models we trained would strongly ``overfit'' to the sensory data and predictive performance (as induced by label binding) would drop during longer, multi-pass trials; overfitting in our experimental settings meant that the synaptic efficacies seemed to saturate to the extreme value boundaries (in our models, this was towards either zero or one, as synaptic weights were constrained to be non-negative and be bounded to the range $[0,1]$) yielding `brittle' or `rigid' synaptic templates that could only match with a few actual sensory samples. 
%While working with the models another limitation that was found is that without being weight dependant the models will overfit and actually drop in performance over long running trials. 
This problematic effect is likely due to weights being clipped at one and the non-scaled weight updates driving all values towards one (the maximal allowed synaptic efficacy), hence why weight-dependency scaling (or soft-bounding) was found to be so important.

Based on our experimental results, we can also see that all of the spiking models get confused by digits that look the same; this we hypothesize is due to the nature of LIF dynamics since, if there is enough overlap between two classes to produce a spike from the hidden layer of LIFs, then one class can appear as a subset of the others. This can generally be tackled with lateral inhibition, as we empirically confirmed heavily in preliminary experimentation, though we remark that the inhibitory-to-excitatory (and excitatory-to-inhibitory) scaling factors used in the first experimental case uses are rather extreme and limit much of the dynamics to, in many cases, a single winner-take-all style of competition. This form of cross-layer competition can be somewhat limiting when attempting to construct deeper, multi-layer neuronal circuits.

The final limitation we find useful to point out is with respect to the LIF neuronal dynamics. Across all of the models tested, if the LIF settings resulted in neuronal layers producing too many, or not enough, spikes, we found that none of the models were able to effectively learn. In addition to this, we found that if the dynamics was not carefully tuned, spikes would appear in waves (or ``volleys'') causing temporal information to be lost and training to ultimately be much harder.

\section{Conclusions}
\label{sec:conclusions}

In this study, we proposed time-integrated spike-timing-dependent plasticity (TI-STDP), a novel generalization of spike-timing centered mathematical formulations of long-term synaptic  plasticity. TI-STDP does not require the maintenance or explicit introduction of auxiliary trace variables as well as any events to induce adjustments made to synaptic efficacies, embodying the core tenets of canonical STDP in a convenient, flexible computational framework. Our experimental results demonstrated TI-STDP's effectiveness as a mechanism for conducting biologically-motivated credit assignment in multi-layer spiking neuronal circuits, further outperforming several key historical forms of STDP with respect to online learning in the context of a pattern classification task. Furthermore, we formulated a simple compositional biophysical model that we demonstrated, crafting an approximate ancestral assembly process, was capable of learning how to synthesize more complex object patterns from simpler, part-like elements extracted from sensory by spiking feature detectors.

\subsection*{Acknowledgements}

This material is based upon work supported by the National Science 
Foundation under Award No. DGE-2125362. Any opinions, findings, 
and conclusions or recommendations expressed in this material are 
those of the author(s) and do not necessarily reflect the views of the 
National Science Foundation.

\bibliographystyle{acm}
\bibliography{ref}

\newpage
\input{appendix}

\end{document}

%% file: appendix.tex
\section*{Supplementary Material}

\begin{figure}[!t]
    \centering
    \begin{subfigure}[b]{0.485\textwidth}
        \includegraphics[width=0.9\textwidth]{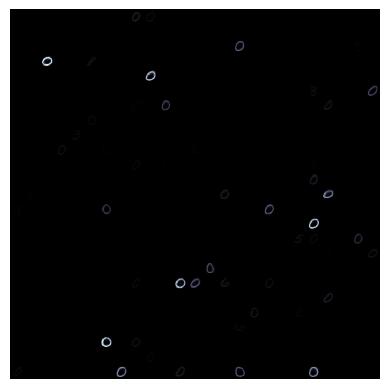} 
    \end{subfigure}
    \begin{subfigure}[b]{0.485\textwidth}
        \includegraphics[width=0.9\textwidth]{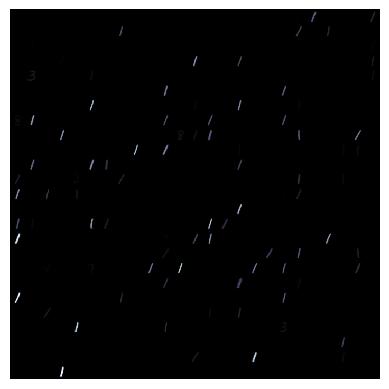} 
    \end{subfigure}\\
    \begin{subfigure}[b]{0.485\textwidth}
        \includegraphics[width=0.9\textwidth]{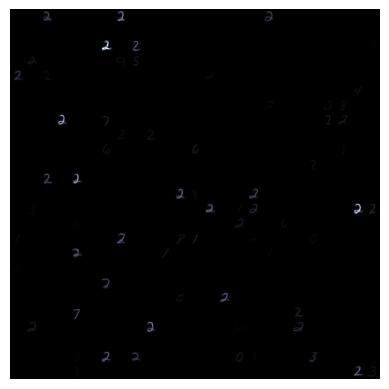} 
    \end{subfigure}
    \begin{subfigure}[b]{0.485\textwidth}
        \includegraphics[width=0.9\textwidth]{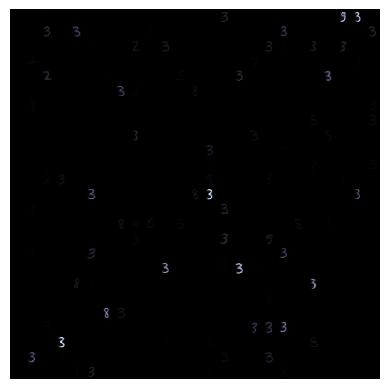} 
    \end{subfigure}\\
    \begin{subfigure}[b]{0.485\textwidth}
        \includegraphics[width=0.9\textwidth]{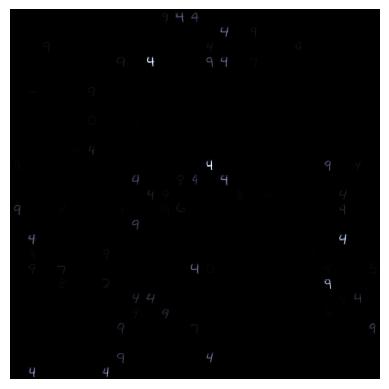} 
    \end{subfigure}
    \begin{subfigure}[b]{0.485\textwidth}
        \includegraphics[width=0.9\textwidth]{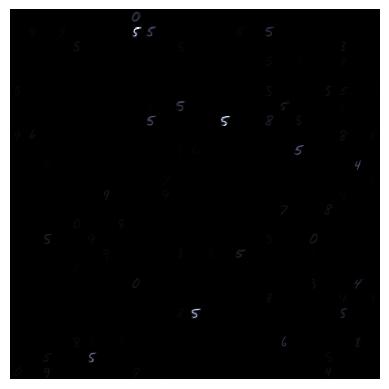} 
    \end{subfigure}
    \caption{\textbf{TI-STDP Acquired Pattern Clusters for Digits $0$ through $5$.} Visualization of pattern clusters that were extracted under different spike-timing-driven algorithm conditions} \label{fig:tistdp_digit_clusters_1to5}
\end{figure}

\begin{figure}[!t]
    \centering
    \begin{subfigure}[b]{0.485\textwidth}
        \includegraphics[width=0.9\textwidth]{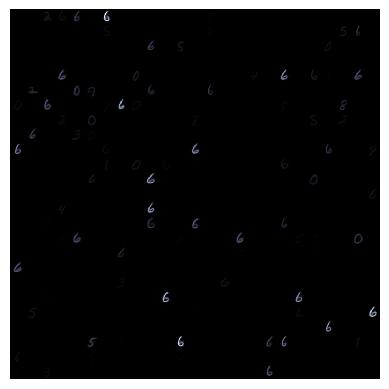} 
    \end{subfigure}
    \begin{subfigure}[b]{0.485\textwidth}
        \includegraphics[width=0.9\textwidth]{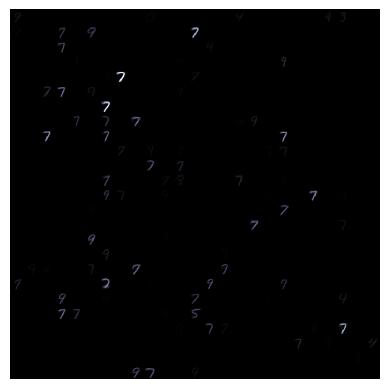} 
    \end{subfigure}\\
    \begin{subfigure}[b]{0.485\textwidth}
        \includegraphics[width=0.9\textwidth]{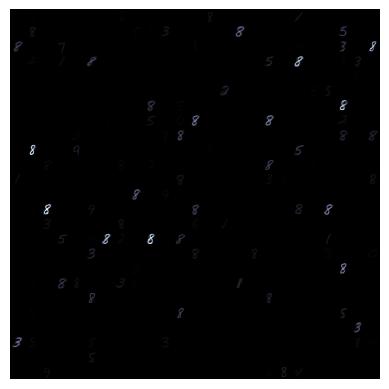} 
    \end{subfigure}
    \begin{subfigure}[b]{0.485\textwidth}
        \includegraphics[width=0.9\textwidth]{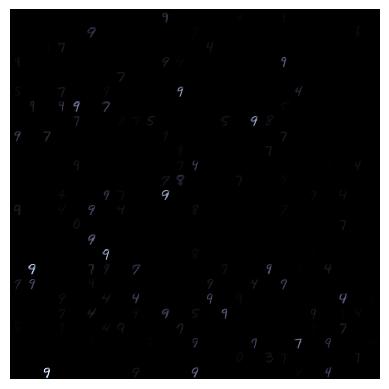} 
    \end{subfigure}
    \caption{\textbf{TI-STDP Acquired Pattern Clusters for Digits $6$ through $9$.} Visualization of pattern clusters that were extracted under different spike-timing-driven algorithm conditions} \label{fig:tistdp_digit_clusters_6to9}
\end{figure}

In this appendix, we present several visualizations of sampled clusters acquired by our multi-level biophysical models trained on full, non-patched digit images. Figures \ref{fig:tistdp_digit_clusters_1to5} and \ref{fig:tistdp_digit_clusters_6to9} showcase at least one representative cluster per each unique digit class/category acquired by the TI-STDP-adapted SNN model. Note that brighter, more white digits indicate more strongly correlated patterns that are associated with any given visualized cluster, while less bright/more faded digits indicate lesser-associated patterns with a given cluster (but still strong enough to be included as part of that cluster which, in some cases, highlights some of the mismatches/mistakes made by the model's self-organizing second layer). 

%% file: main.bbl
\begin{thebibliography}{10}

\bibitem{bellec2020solution}
{\sc Bellec, G., Scherr, F., Subramoney, A., Hajek, E., Salaj, D., Legenstein, R., and Maass, W.}
\newblock A solution to the learning dilemma for recurrent networks of spiking neurons.
\newblock {\em Nature communications 11}, 1 (2020), 3625.

\bibitem{bi1998synaptic}
{\sc Bi, G.-q., and Poo, M.-m.}
\newblock Synaptic modifications in cultured hippocampal neurons: dependence on spike timing, synaptic strength, and postsynaptic cell type.
\newblock {\em Journal of neuroscience 18}, 24 (1998), 10464--10472.

\bibitem{bi2001synaptic}
{\sc Bi, G.-q., and Poo, M.-m.}
\newblock Synaptic modification by correlated activity: Hebb's postulate revisited.
\newblock {\em Annual review of neuroscience 24}, 1 (2001), 139--166.

\bibitem{carafoli1987intracellular}
{\sc Carafoli, E.}
\newblock Intracellular calcium homeostasis.
\newblock {\em Annual review of biochemistry 56}, 1 (1987), 395--433.

\bibitem{crick1989recent}
{\sc Crick, F.}
\newblock The recent excitement about neural networks.
\newblock {\em Nature 337}, 6203 (1989), 129--132.

\bibitem{diehl2015unsupervised}
{\sc Diehl, P.~U., and Cook, M.}
\newblock Unsupervised learning of digit recognition using spike-timing-dependent plasticity.
\newblock {\em Frontiers in computational neuroscience 9\/} (2015), 99.

\bibitem{draghici2000neural}
{\sc Draghici, S.}
\newblock Neural networks in analog hardware—design and implementation issues.
\newblock {\em International journal of neural systems 10}, 01 (2000), 19--42.

\bibitem{ferre2018unsupervised}
{\sc Ferr{\'e}, P., Mamalet, F., and Thorpe, S.~J.}
\newblock Unsupervised feature learning with winner-takes-all based stdp.
\newblock {\em Frontiers in computational neuroscience 12\/} (2018), 24.

\bibitem{fisher2022recursive}
{\sc Fisher, A., and Rao, R.~P.}
\newblock Recursive neural programs: Variational learning of image grammars and part-whole hierarchies.
\newblock {\em arXiv preprint arXiv:2206.08462\/} (2022).

\bibitem{furber2016large}
{\sc Furber, S.}
\newblock Large-scale neuromorphic computing systems.
\newblock {\em Journal of neural engineering 13}, 5 (2016), 051001.

\bibitem{gerstner1996neuronal}
{\sc Gerstner, W., Kempter, R., Van~Hemmen, J.~L., and Wagner, H.}
\newblock A neuronal learning rule for sub-millisecond temporal coding.
\newblock {\em Nature 383}, 6595 (1996), 76--78.

\bibitem{grossberg1987resonance}
{\sc Grossberg, S.}
\newblock Competitive learning: From interactive activation to adaptive resonance.
\newblock {\em Cognitive Science 11}, 1 (1987), 23 -- 63.

\bibitem{gupta2009hebbian}
{\sc Gupta, A., and Long, L.~N.}
\newblock Hebbian learning with winner take all for spiking neural networks.
\newblock In {\em 2009 International Joint Conference on Neural Networks\/} (2009), IEEE, pp.~1054--1060.

\bibitem{hebb1949organization}
{\sc Hebb, D.~O.}
\newblock {\em The organization of behavior: A neuropsychological theory}.
\newblock New York: Wiley, 1949.

\bibitem{hinton1979some}
{\sc Hinton, G.}
\newblock Some demonstrations of the effects of structural descriptions in mental imagery.
\newblock {\em Cognitive Science 3}, 3 (1979), 231--250.

\bibitem{huang2004glutamate}
{\sc Huang, Y.~H., and Bergles, D.~E.}
\newblock Glutamate transporters bring competition to the synapse.
\newblock {\em Current opinion in neurobiology 14}, 3 (2004), 346--352.

\bibitem{karmarkar2002mechanisms}
{\sc Karmarkar, U.~R., Najarian, M.~T., and Buonomano, D.~V.}
\newblock Mechanisms and significance of spike-timing dependent plasticity.
\newblock {\em Biological cybernetics 87}, 5 (2002), 373--382.

\bibitem{kempter1999hebbian}
{\sc Kempter, R., Gerstner, W., and Van~Hemmen, J.~L.}
\newblock Hebbian learning and spiking neurons.
\newblock {\em Physical Review E 59}, 4 (1999), 4498.

\bibitem{lake2015human}
{\sc Lake, B.~M., Salakhutdinov, R., and Tenenbaum, J.~B.}
\newblock Human-level concept learning through probabilistic program induction.
\newblock {\em Science 350}, 6266 (2015), 1332--1338.

\bibitem{lecun1998mnist}
{\sc LeCun, Y.}
\newblock The mnist database of handwritten digits.
\newblock {\em http://yann.lecun.com/exdb/mnist/\/} (1998).

\bibitem{maass1997networks}
{\sc Maass, W.}
\newblock Networks of spiking neurons: the third generation of neural network models.
\newblock {\em Neural networks 10}, 9 (1997), 1659--1671.

\bibitem{markram2011history}
{\sc Markram, H., Gerstner, W., and Sj{\"o}str{\"o}m, P.~J.}
\newblock A history of spike-timing-dependent plasticity.
\newblock {\em Frontiers in synaptic neuroscience 3\/} (2011), 4.

\bibitem{markram1997physiology}
{\sc Markram, H., L{\"u}bke, J., Frotscher, M., Roth, A., and Sakmann, B.}
\newblock Physiology and anatomy of synaptic connections between thick tufted pyramidal neurones in the developing rat neocortex.
\newblock {\em The Journal of physiology 500}, 2 (1997), 409--440.

\bibitem{markram1997regulation}
{\sc Markram, H., L{\"u}bke, J., Frotscher, M., and Sakmann, B.}
\newblock Regulation of synaptic efficacy by coincidence of postsynaptic aps and epsps.
\newblock {\em Science 275}, 5297 (1997), 213--215.

\bibitem{massa2020efficient}
{\sc Massa, R., Marchisio, A., Martina, M., and Shafique, M.}
\newblock An efficient spiking neural network for recognizing gestures with a dvs camera on the loihi neuromorphic processor.
\newblock In {\em 2020 International Joint Conference on Neural Networks (IJCNN)\/} (2020), IEEE, pp.~1--9.

\bibitem{mead1990neuromorphic}
{\sc Mead, C.}
\newblock Neuromorphic electronic systems.
\newblock {\em Proceedings of the IEEE 78}, 10 (1990), 1629--1636.

\bibitem{ororbia2023learning}
{\sc Ororbia, A.}
\newblock Contrastive-signal-dependent plasticity: Forward-forward learning of spiking neural systems.
\newblock {\em arXiv preprint arXiv:2303.18187\/} (2023).

\bibitem{ororbia2023spiking}
{\sc Ororbia, A.}
\newblock Spiking neural predictive coding for continually learning from data streams.
\newblock {\em Neurocomputing 544\/} (2023), 126292.

\bibitem{ororbia2024review}
{\sc Ororbia, A., Mali, A., Kohan, A., Millidge, B., and Salvatori, T.}
\newblock A review of neuroscience-inspired machine learning.
\newblock {\em arXiv preprint arXiv:2403.18929\/} (2024).

\bibitem{ororbia2023brain}
{\sc Ororbia, A.~G.}
\newblock Brain-inspired machine intelligence: A survey of neurobiologically-plausible credit assignment.
\newblock {\em arXiv preprint arXiv:2312.09257\/} (2023).

\bibitem{robertson1991neuropsychological}
{\sc Robertson, L.~C., and Lamb, M.~R.}
\newblock Neuropsychological contributions to theories of part/whole organization.
\newblock {\em Cognitive psychology 23}, 2 (1991), 299--330.

\bibitem{samadi2017deep}
{\sc Samadi, A., Lillicrap, T.~P., and Tweed, D.~B.}
\newblock Deep learning with dynamic spiking neurons and fixed feedback weights.
\newblock {\em Neural computation 29}, 3 (2017), 578--602.

\bibitem{stein1965theoretical}
{\sc Stein, R.~B.}
\newblock A theoretical analysis of neuronal variability.
\newblock {\em Biophysical journal 5}, 2 (1965), 173--194.

\bibitem{tavanaei2018representation}
{\sc Tavanaei, A., Masquelier, T., and Maida, A.}
\newblock Representation learning using event-based stdp.
\newblock {\em Neural Networks 105\/} (2018), 294--303.

\bibitem{van2008visualizing}
{\sc Van~der Maaten, L., and Hinton, G.}
\newblock Visualizing data using t-sne.
\newblock {\em Journal of machine learning research (JMLR) 9\/} (2008), pp. 2579--2605.

\bibitem{zhang2021optical}
{\sc Zhang, H., Gu, M., Jiang, X., Thompson, J., Cai, H., Paesani, S., Santagati, R., Laing, A., Zhang, Y., Yung, M.-H., et~al.}
\newblock An optical neural chip for implementing complex-valued neural network.
\newblock {\em Nature communications 12}, 1 (2021), 457.

\end{thebibliography}
